\documentclass[12pt]{iopart}
\usepackage[latin1]{inputenc}
\usepackage{iopams}
\usepackage[english]{babel}
\newtheorem{thm}{Th\'eor\`eme}[section]
\newtheorem{cor}[thm]{Corollaire}
\newtheorem{lem}[thm]{Lemme}
\newtheorem{pro}[thm]{Proposition}
\newtheorem{dfn}[thm]{D\'efinition}
\newtheorem{rmq}[thm]{Remark}
\newtheorem{expl}[thm]{Exemple}

\eqnobysec

\newcommand{\R}{\mathbb R}
\newcommand{\C}{\mathbb C}
\newcommand{\N}{\mathbb N}

\newcommand{\1}{1 \! \! {\rm I}}

\newcommand{\beq}{\begin{eqnarray}}
\newcommand{\eeq}{\end{eqnarray}}
\newcommand{\bpro}{\begin{pro}}
\newcommand{\epro}{\end{pro}}
\newcommand{\blem}{\begin{lem}}
\newcommand{\elem}{\end{lem}}
\newcommand{\bdfn}{\begin{dfn}}
\newcommand{\edfn}{\end{dfn}}
\newcommand{\bcor}{\begin{cor}}
\newcommand{\ecor}{\end{cor}}
\newcommand{\bthm}{\begin{thm}}
\newcommand{\ethm}{\end{thm}}
\newcommand{\bex}{\begin{expl}}
\newcommand{\eex}{\end{expl}}
\newcommand{\brmq}{\begin{rmq}}
\newcommand{\ermq}{\end{rmq}}
\newcommand{\benum}{\begin{enumerate}}
\newcommand{\eenum}{\end{enumerate}}
\newcommand{\bitem}{\begin{itemize}}
\newcommand{\eitem}{\end{itemize}}
\begin{document}

\title{{Vector coherent
states  for nanoparticle Hamiltonians}}

\author{Isiaka Aremua$^1$ and Mahouton Norbert Hounkonnou$^2$\footnote{Corresponding author:
norbert.hounkonnou@cipma.uac.bj  with copy to hounkonnou@yahoo.fr.} }

\address{$^1$ Institut de Math\'ematiques et de Sciences Physiques (IMSP)\\
University of Abomey-Calavi\\
01 B.P. 613 Porto-Novo, Republic of Benin}

\address{$^2$ International Chair of Mathematical Physics
and Applications \\
(ICMPA-UNESCO Chair),
University of Abomey-Calavi\\
072 B.P. 50 Cotonou, Republic of Benin}

\ead{iaremua@imsp-uac.org and norbert.hounkonnou@cipma.uac.bj}

\begin{abstract}
The first part of this work deals  with a
formalism of vector coherent states construction 
for  a system of $M$ Fermi-type  modes  associated
with $N$ bosonic modes. Then follows   a  generalization to
a Hamiltonian describing the translational motion of the center of
mass of a nanoparticle. The latter gives rise to a new mechanism for the
electronic energy relaxation in nanocrystals,  intensively
studied today in condensed matter physics. Finite
degeneracies  of the involved Hamiltonian systems are also investigated.
The defined vector coherent states satisfy relevant mathematical
properties of continuity, resolution of identity, temporal stability
and action identity.
\end{abstract}



\section{Introduction}
Nanoscience has exploded in the last decade, primarily as the result
of the development of new tools that have made the characterization
and manipulation of nanostructures practical, and also as a result
of new methods for preparation of these structures. Many
technological applications including miniaturization in
microelectronics devices yield research activities on electronic
properties of new nanoscale structure. In at least one dimension,
when the size of an object becomes comparable to the characteristic
length of an electronic quantum effect, its physical manifestation
is modified: interference and coherence effects, quantum tunneling,
discrete energy level, collective effects, electronic transport,
role of interactions, coupling to the environment must be taken into
account. All these  phenomena occurring in quantum electronics lie at
the heart of today scientific activities to which intense efforts
are devoted due to their numerous applications in various domains
(\cite{joben}, \cite{komi} and references therein). Indeed, the
research activities in   quantum nanostructures are useful for the development of the
fields of  quantum information (quantum computers), spin electronic
(spintronics), molecular electronics and nanophotonics. In the field
of spintronics, the understanding of spin-related phenomena in
magnetic or non-magnetic doped nanostructures is of paramount
importance for using spin as information vector. Spin dynamics is
studied in quantum confined nanostructures and ferromagnetic
semiconductor layers by pump-probe techniques. Magnetic domain
structure and domain wall motion are investigated in ferromagnetic
semiconductors (also in  superconductors) using magneto-optical Kerr
microscopy. Electronic excitations in two-dimensional $(2D)$ and
one-dimensional nanostructures are investigated by means of
electronic Raman spectroscopy: collective and individual spin
polarized excitations of a $(2D)$ electron gas,  electronic
excitations in quantum wells and wires. Moreover, a new development
in near-field imaging \cite{garcia} now makes it possible to map
vector fields on the nanoscale as never before. Lee and co-workers
\cite{lee}  recently reported an experimental technique that can
capture and map the vectorial nature of the electric fields down to
the nanoscale. This could lead to important applications in physics
and biochemistry. Indeed, in-depth knowledge of the electric field
vector on the nanoscale could help in the design of miniaturized
optical components that may replace their electronic counterparts in
the future. In addition, near-field vector imaging is also important
in biosensing applications because the interaction between light and
biological molecules strongly depends on the orientation of the
electric field \cite {garcia}. By extracting this information we can
uncover new effects at play.

In the theory of electron-phonon dynamics in
insulating nanoparticles context, the electron-phonon dynamics in an
ensemble of nearly isolated nanoparticles, the vibrational dynamics
of nanometer-scale
 semiconducting  and insulating are probed by localized impurity states (\cite{simon-geller}).
 Nanoparticles are  modeled there as  electronic two-level systems coupled to  single vibrational modes.

Despite all these exciting explorations, to the best of our
knowledge, not much has been done in the investigation of vector
coherent states for nanoparticle systems. This work aims at filling
this gap.

 Coherent states originally introduced by Schr\"{o}dinger
(\cite{sch}) in the context of a harmonic oscillator and later
popularized by Glauber and Klauder for the description of coherent
light are generally acknowledged  to provide a close connection
between classical and quantum formulations for a given system.
Indeed, they are useful for the  phenomenological description of a
nonlinear optical environment and nonlinear interacting systems in
quantum optics.  The concept of vector coherent states (VCSs) has
been introduced in mathematical literature in fifties (see
\cite{bott} and references therein) in the study of induced
representations of groups, constructed using vector bundles over
homogeneous spaces and, while in the physical literature the idea
was born  with their formulation  in connection with the use of the
symplectic group for describing collective models of nuclei
(\cite{ali-englis-gazeau} and references therein). Ever since,  they
have been widely used in a variety of symmetry problems in quantum
mechanics in general, and in particular  for quantum optical models
(\cite{berube-hussin-nieto}, \cite{daoud-hussin},
\cite{ali-englis-gazeau}), as well as for the study of spectra of
two-level atomic systems  in electromagnetic fields, for instance
 in the famous Jaynes-Cummings model (\cite{jcum}). In
previous works, in the context of quantum optics, a new formulation
of VCSs for nonlinear spin-orbit Hamiltonian model \cite{joben}  in
terms of the matrix eigenvalue problem for generalized annihilation
operators has been provided. Let us cite also the VCSs of
Gazeau-Klauder type  \cite{ab}, constructed
 for a Hamiltonian  describing the interaction between a single mode, $(a,a^{\dag})$ of the
radiation field with the frequency $\omega$ and $2$ fermion levels
with energies  $\epsilon_{1}, \epsilon_{2}, $ and creation and
annihilation operators $c^{\dag}_{j},c_{j}, j=1,2.$

This paper addresses investigations on
 general construction of  VCSs
for nanoparticle systems which are today broadly used in the domain
of condensed matter physics and include the physical system studied
in \cite{ab}  as a particular case. In section 2,  VCSs for the
electron-phonon dynamics are built. Then, in section 3, we provide
with a  generalization of examined model and construct corresponding
coherent states. Finally we end with some concluding remarks in section 4.
Furthermore, in order to make clearer the comprehension of the main results of the paper,
 we present in the  appendices a detailed computation
of some useful relations.

\section{Vector coherent states  for electron-phonon dynamics}
In this section, we deal with the construction of vector coherent
states for the  model describing the electron-phonon dynamics. This
model consists of   the following Hamiltonian given (with $\hbar =
1$) by \beq{\label{hamil} } H = \sum_{i=1}^{N}\omega_i a^{\dag}_i
a_i + \sum_{j=1}^{M}\epsilon_{j}c^{\dag}_jc_j +
\sum_{i=1}^{N}\sum_{l=1}^{M}g_l c^{\dag}_l c_l (a^{\dag}_i + a_i)
\eeq where the following commutation rules
\beq{\label{commuta}}
 && \fl [a_{_{i}}, a^{^{\dag}}_{k}] = \delta_{ik}\1 \qquad
\{c^{\dag}_{j}, c_{l}\} = \delta_{jl}\1 \cr
&& \fl [a_{i}, a_{k}] = 0 \qquad
[a^{\dag}_{i}, a^{\dag}_{k}]=0 \qquad   \{c_{j}, c_{l}\} = 0 \qquad
\{c^{\dag}_{j}, c^{\dag}_{l}\}=0 \qquad    \eeq
with $1\leq i, k \leq N,$  $1 \leq j, l \leq
M$, hold. The set  $\{a_{i},a^{\dag}_{i},N_{i} = a^{\dag}_{i}a_{i}\}$
for each $i \, (1\leq i \leq N)$ span the ordinary Fock-Heisenberg
oscillator algebra.

The first term of this Hamiltonian describes the internal
vibrational dynamics of a nanoparticle; the $ \omega_{i}$ are
frequencies of the internal modes, and $a^{\dag}_{i}$ and $a_{i}$
are the corresponding phonon creation and annihilation operators.
The second term  is the Hamiltonian for a noninteracting $M$-level
system while the third term stands for the ordinary leading-order interaction
between the $M$-level system and the internal vibrational modes. The
constants $g_{l} (l=1,2,\cdots, M)$  are the electron-phonon
interaction energy. This system generalizes a previous work
\cite{ab} on a specific physical Hamiltonian    of a single photon
mode with the frequency $\omega$ interacting with a pair of
fermions, for which  vector coherent states of Gazeau-Klauder type
have been explicitly constructed, satisfying the four requirements
of continuity, temporal stability, resolution of the identity and
action identity \cite{klauder2}.  The same model,  describing an
interaction between a single mode, $(a, a^\dag)$, of the radiation
field with two Fermi type modes, has been also
investigated by Simon and Geller who outlined  some physical
aspects of the ensemble-averaged excited-state population dynamics
 \cite{simon-geller} and showed its   relevance in the
study of electron-phonon dynamics in an ensemble of nearly isolated
nanoparticles, in the context of quantum
effects in condensed matter systems. The vibrational spectrum of the
nanoparticle is here provided by the localized electronic impurity
states in doped nanocrystal
  \cite{hong}, \cite{yang}. The impurity states can be used to probe the
  energy relaxation by phonon emission.

    The generalized Hamiltonian (\ref{hamil})
  considered in this work describes a nanoparticle modeled as a
  system  of
  $M$ Fermi
  type modes interacting with  $N$ vibrational radiation modes
  (phonons) with frequencies $\omega_{i}, 1 \leq i \leq N.$
The eigenvector of the Hamiltonian $H$ can be written as \beq
\varphi =  \Phi_1 \otimes \Phi_2 \otimes \cdots \otimes \Phi_N \otimes \Psi_{[{\bf k}]} \qquad [{\bf k}]= k_1,k_2,\dots, k_M \eeq
where
$\Phi_{l}, l = 1,2,\dots, N$ and $
\Psi_{[{\bf k}]}$  are  the bosonic and fermion states, respectively,
with  $k_1, k_2, \dots, k_M = 0, 1.$ For all  $k_j, 1 \leq j \leq M,
$ the eigenvalue problem can be then defined as \beq{\label{a}}
H\varphi=\left[\sum_{i=1}^{N}\omega_i a^{\dag}_i a_i +
\sum_{i=1}^{N}\sum_{j=1}^{M}\delta_{k_j,1}(\epsilon_j +
g_j(a^{\dag}_i + a_i))\right]\varphi:=E\varphi \eeq $E$ being the
eigenenergy  of the system.
 As a matter of appropriate analysis, let us consider the following self-adjoint
operator $B_{[{\bf k}]_{1,2,\dots, N}}$ given by  \cite{ab} \beq B_{[{\bf
k}]_{1,2,\dots, N}} = \sum_{l=1}^{N} B_{[{\bf k}]_l} \qquad B_{[{\bf
k}]_l} = \omega_l A^{\dag}_{[{\bf k}]_l}A_{[{\bf k}]_l} +
\frac{\epsilon_{[{\bf k}]}}{N} - \frac{g^2_{[{\bf k}]}}{\omega_l}
\eeq with
\beq{\label{self}}
[A_{[{\bf k}]_l}, A^{\dag}_{[{\bf k}]_l}] =\1 \qquad
 A_{[{\bf k}]_l} = a_l + \frac{g_{[{\bf k}]}}{\omega_l} \qquad l = 1, 2 ,\dots
 N
\eeq where we have introduced  the quantities $g_{[\bf k]}$ and
$\epsilon_{[{\bf k}]}$, with the dimension of energy (taking $\hbar =
1$), as \beq g_{[\bf k]}:=\sum_{k_{j} \in[{\bf
k}] }k_{j}g_{j}\qquad \epsilon_{[{\bf k}]}:=\sum_{k_{j} \in[{\bf
k}] }k_{j}\epsilon_{j} \qquad 1\leq j \leq M  \qquad [{\bf k}] \in \Gamma \eeq
where
\beq{\label{set00}}
\Gamma = \{(0,0,\cdots, 0,0), (1,0, \cdots, 0,0), \dots, (1,1\cdots, 1,1) \}.
\eeq

The operators
$A_{[{\bf k}]_l}$ are such that \beq A_{[{\bf k}]_l} =
e^{\imath\sqrt{2}\frac{g_{[{\bf k}]}}{\omega_l}P_l} a_l e^{-\imath
\sqrt{2}\frac{g_{[{\bf k}]}}{\omega_l}P_l} \qquad P_l =
\frac{a_l-a^{\dag}_l}{\imath \sqrt{2}} \qquad l = 1, 2, \cdots, N \eeq
where, for all $l$, the commutation relation \beq
[a_{l},\sqrt{2}P_{l}] =  \imath \1 \eeq is satisfied. Thus, for all $l,
l=1, 2, \cdots, N$, the eigenstates $|\Phi^{[{\bf k}]}_{n_l}\rangle$
of  $B_{[{\bf k}]_l}$ are given by \beq |\Phi^{[{\bf
k}]}_{n_l}\rangle =  e^{\imath \sqrt{2}\frac{g_{[{\bf
k}]}}{\omega_l}P_l}|n_l\rangle = \frac{(A^{\dag}_{[{\bf
k}]_l})^{n_l}} {\sqrt{n_l !}}|\Phi^{[{\bf k}]}_0\rangle_l \eeq
where $ |\Phi^{[{\bf k}]}_0 \rangle_l = e^{\imath
\sqrt{2}\frac{g_{[{\bf k}]}}{\omega_l}P_l}|0 \rangle$ with the
corresponding  eigenvalues  \beq E^{[{\bf k}]}_{n_l} = \omega_l n_l
+ \frac{\epsilon_{[{\bf k}]}}{N} - \frac{g^2_{[{\bf k}]}}{\omega_l}.
\eeq Therefore, the eigenstates of $B_{[{\bf k}]_{1,2,\dots, N}}$
can be expressed as \beq{\label{c}} |\Phi^{[{\bf k}]}_{n_1,n_2,\cdots, n_N}\rangle
&=& |\Phi^{[{\bf k}]}_{n_1}\rangle \otimes |\Phi^{[{\bf
k}]}_{n_2}\rangle \otimes\cdots \otimes |\Phi^{[{\bf
k}]}_{n_N}\rangle \cr
                                      &=&\left( \bigotimes_{l=1}^{N}
\frac{(A^{\dag}_{[{\bf k}]_l})^{n_l}}{\sqrt{n_l
!}}e^{\imath\sqrt{2}\frac{g_{[{\bf
k}]}}{\omega_l}P_l}\right)(|0\rangle \otimes |0\rangle \otimes
\cdots \otimes |0\rangle) \eeq $n_1, n_2, \dots, n_N = 0, 1, 2,
\dots,$ associated with
 the eigenvalues
 \beq{\label{d}} 
E^{[{\bf
k}]}_{n_1,n_2,\dots, n_N} = \sum_{l=1}^{N}\omega_l n_l +
\epsilon_{[{\bf k}]} - g^2_{[{\bf k}]}\sum_{l=1}^{N}
\frac{1}{\omega_l}
\eeq
where $l = 1, 2, \dots, N, \; n_l= 0, 1, 2, \dots, \; [{\bf k}] \in \Gamma$. 

 Physically,  $ {[{\bf k}]}, k_{1}, k_{2}, \dots, k_{M} = 0,1$,
can be interpreted as the two possible arrangements of fermion
particles corresponding to ferromagnetism or antiferromagnetism
order, i.e.  $[{\bf k}] = (0,0,\dots, 0, \dots, 0)$ corresponds to the
situation where  all  the $M$ spins are down while $[{\bf k}] =
(1,1,\dots, 1, \dots, 1)$ represents the case when they are all up; the
fermion states $\Psi_{[{\bf k}]}$ correspond to spin waves, also
called {\it magnons} by analogy to phonons \cite{kittel} which
constitute elementary excitations of the $M$ spins system,  generated
by  the interaction of vibrational modes (or bosonic modes) of
frequencies $\omega_{i}, i = 1, 2, \dots, N, $  with the $M$
fermion levels.  The eigenstates of this Hamiltonian,  defined
as the tensor product of the bosonic and fermion states,  can be
thus written as
\beq{\label{e}} 
\fl \varphi^{[{\bf k}]}_{n_1,n_2,\dots,
n_N} =  \Phi^{[{\bf k}]}_{n_1,n_2,\dots, n_N} \otimes \Psi_{[{\bf k}]} \qquad
\Psi_{[{\bf k}]} = (c^{\dag}_1)^{k_1}(c^{\dag}_2)^{k_2}\cdots
(c^{\dag}_j)^{k_j} \cdots (c^{\dag}_{M})^{k_M} \Psi_{(0,0,\dots, 0)}
\eeq where $\Psi_{(0,0,\dots, 0)}$ is the fermion vacuum,  with $ k_1,
k_2, \dots, k_M = 0, 1$, corresponding to  the associated eigenvalues
 (\ref{d}).

\subsection{Vector coherent states for the nondegenerate Hamiltonian}
Let us consider  the Hamiltonian $H$ with a discrete positive
spectrum and  eigenvectors spanning a Hilbert space $\mathfrak
H.$ With $[{\bf k}]=(k_{1},k_{2},\cdots, k_{M})$ where  $k_{1},k_{2},
\cdots ,k_{M}=0,1, $ these vectors can be grouped into $2^{M}$
families. Then $\mathfrak H = \bigoplus_{[{\bf k}] \in \Gamma}
\mathfrak H_{[{\bf k}]}$,
and $\mathfrak H_{[{\bf k}]}$ is the
subspace of $\mathfrak H$ spanned by the vectors $\varphi^{[{\bf
k}]}_{n_1,n_2,\dots, n_N}$ for every $[{\bf k}]$.
This subspace remains
stable under the action of the projection operator denoted  by
$\mathbb P_{[{\bf k}]}.$

Furthermore, assume that all the $\omega_l$ are different, for all $
l, 1\leq l \leq N$, and $E^{[{\bf k}]}_{n_l} - E^{[{\bf k}]}_{0} =
\omega_l n_l$.
In particular  for any two given bosonic modes $\omega_i, \omega_j$, $i \neq
j$, let us take the quantity $\omega_i/\omega_j \not\in  \mathbb Q$. Then,  for two shifted eigenvalues $E^{[{\bf k}]}_{n_i} - E^{[{\bf k}]}_{0} =
\omega_i n_i$ and $E^{[{\bf k}]}_{n_j} - E^{[{\bf k}]}_{0} =
\omega_j n_j$, writing the energy spectrum in the form
\beq
\mathcal E^{[{\bf k}]}_{n_{i},n_{j}} \equiv \frac{n_{i}\omega_{i} + n_{j}\omega_{j}}{\omega_{j}} = n_{i}\frac{\omega_i}{\omega_j} + n_{j}
\eeq
one infers that there is no information about the degeneracy. For $\omega_i/\omega_j  = p/q \in \mathbb Q$, then it is possible and we get
\beq
\mathcal E^{[{\bf k}]}_{n_{i},n_{j}} = \mathcal E^{[{\bf k}]}_{n'_{i},n'_{j}} \quad \mbox{if} \quad \frac{p}{q} 
= \frac{n_{j} - n'_{j}}{n'_{i} - n_{i}}.
\eeq
 Setting \beq{\label{lu}}
n_l :=  \epsilon_{n_{l}}= \frac{E^{[{\bf k}]}_{n_l} - E^{[{\bf k}]}_{0}}{\omega_l} \eeq
leads to a sequence of dimensionless quantities  $
(\epsilon_{n_l})_{n_l \in \N}$ satisfying the inequalities $
0=\epsilon_0 <\epsilon_1 <\epsilon_2 <\dots.$

We define the unitary map $V : \mathfrak H \rightarrow \C^{2^{M}} \otimes \hat \mathfrak H $
such that  $V\varphi^{[{\bf k}]}_{[\bf n]} = \Psi_{[{\bf k}]} \otimes \Phi^{[{\bf k}]}_{[\bf n]},$  where $\hat \mathfrak H$ is spanned
by the bosonic states $\left\{|\Phi^{[{\bf k}]}_{[\bf n]}\rangle, [{\bf k}] \in \Gamma \right\}_{[{\bf n}]=0}^{\infty}$. Formally, the operator  
$V$ can be written as
\beq
V = \sum_{[{\bf k}] \in \Gamma}\sum_{{[\bf n]} = 0}^{\infty}|\Psi_{[{\bf k}]} \otimes \Phi^{[{\bf k}]}_{[\bf n]}\rangle \langle
\varphi^{[{\bf k}]}_{[\bf n]}|.
\eeq
Diagonalizing $H$ to obtain  $H_{D} =VHV^{\dag}$ as follows:
\beq
\fl H_{D} =VHV^{\dag} =  \left(
\begin{array}{ccccccc}
\tilde H_{(0,\cdots,0,0)} & 0 & \cdots & \cdots & 0  & 0\\
0 & \tilde H_{(1,\cdots,0,0)} &  0 & \cdots & \cdots & \vdots \\
\vdots & 0 & \cdots & 0 & \cdots &\vdots\\
0 & \cdots & 0 & \cdots& 0 & \vdots \\
\vdots & \cdots & \cdots & 0 & \cdots & 0 \\
0 & \cdots & 0 & \cdots & 0 & \tilde H_{(1,\cdots,1,1)} \\
\end{array}
\right)
\eeq
\beq{\label{eq}}
\tilde{H}_{[{\bf k}]}|\Psi_{[{\bf k}]} \otimes \Phi^{[{\bf k}]}_{[\bf n]}\rangle  = E^{[{\bf k}]}_{[\bf n]}|\Psi_{[{\bf k}]} 
\otimes \Phi^{[{\bf k}]}_{[\bf n]}\rangle.
\eeq
The Hamiltonian $H_{D}$ is expressed in terms of $2^{M}$ self-adjoint operators,
 $\tilde{H}_{(0,0,\dots,0,\dots, 0,0)}$, $\tilde{H}_{(1,0,\dots,0,\dots, 0,0)}$,  $\dots, \tilde{H}_{(1,1,\dots,1,\dots,1,1)}$ such  
that each operator, 
being an infinite diagonal matrix with eigenvalues $E^{[{\bf k}]}_{[{\bf n}]}$, acts on  each subspace of 
$\C^{2^{M}} \otimes \hat {\mathfrak H}$ as given
 in (\ref{eq}), with $\tilde{H}_{[{\bf k}]} = VH_{[{\bf k}]}V^{\dag}$, where  $\tilde{H}_{[{\bf k}]}|\Psi_{[{\bf k}]} 
\otimes \Phi^{[{\bf k}]}_{[\bf n]}\rangle  =:H_{[{\bf k}]}| \Phi^{[{\bf k}]}_{[\bf n]} \otimes \Psi_{[{\bf k}]}\rangle.$

Given $l, 1\leq l \leq
N,$ we define the  coherent states  $ |J_{[{\bf
k}]_l},\gamma_{[{\bf k}]_l}\rangle$,  where $ J_{[{\bf k}]_l} > 0$
and $\gamma_{[{\bf k}]_l} \in [0, 2\pi)$ (the variable $J_{[{\bf k}]_l}$
being generally identified with the classical action and
$\gamma_{[{\bf k}]_l}$ with the conjugate angle), as follows: \beq
|J_{[{\bf k}]_l},\gamma_{[{\bf k}]_l}\rangle = (\mathcal N(J_{[{\bf k}]_l}))^{-1/2}\sum_{n_l=0}^{\infty}\frac{J^{n_l/2}_{[{\bf
k}]_l}e^{-\imath \gamma_{[{\bf k}]_l} n_l}}{\sqrt{n_l
!}}|\Psi_{[{\bf k}]}\rangle \otimes |\Phi^{[{\bf k}]}_{n_l}\rangle
\eeq where the vectors $\Psi_{[{\bf k}]}$ form the canonical basis
of $\C^{2^{M}},$ \beq{\label{base}}  && \fl \Psi_{(0,0,\dots,0,\dots0,0)} =
\left(
                          \begin{array}{c}
                           1  \\
                           0   \\
                           \vdots \\
                           0      \\
                           \vdots  \\
                           \vdots     \\
                           0
                          \end{array}
                           \right) \qquad \Psi_{(1,0,\dots,0,\dots,0,0)} = \left(
                          \begin{array}{c}
                            0 \\
                            1  \\
                            0   \\
                           \vdots \\
                           0      \\
                           \vdots   \\
                           0
                          \end{array}
                           \right) \qquad \cdots \quad \Psi_{[{\bf k}]} = \left(
                          \begin{array}{c}
                           0 \\
                           \vdots  \\
                           0 \\
                           1     \\
                           0     \\
                           \vdots  \\
                           0
                          \end{array}
                           \right) \cr
&& \fl \cdots \qquad  \Psi_{(0,1,\dots,1,\dots,1,1)} = \left(
                          \begin{array}{c}
                           0 \\
                           \vdots \\
                           0      \\
                           \vdots  \\
                           0      \\
                           1       \\
                           0
                          \end{array}
                           \right) \qquad \Psi_{(1,1,\dots,1,\dots,1,1)} = \left(
                          \begin{array}{c}
                           0 \\
                           0   \\
                           \vdots \\
                           0      \\
                           \vdots  \\
                           0       \\
                           1
                          \end{array}
                           \right).
\eeq It is easily checked that each family of coherent states
$|J_{[{\bf k}]_l},\gamma_{[{\bf k}]_l}\rangle, \,  l = 1,2,\dots, N$
satisfies the properties of continuity,  temporal stability, action
identity and  resolution of the identity.

Set by definition
\beq{\label{silk}}
\fl {\bf{J}}^{{\bf{n}}/2}_{[\bf{k}]} := \prod_{l=1}^{N} J^{n_l/2}_{[{\bf k}]_l} \qquad 
{\boldsymbol{\gamma}}_{[\bf{k}]} :=
                              \left(
                              \begin{array}{c}
                               \gamma_{[{\bf k}]_1} \\
                              \gamma_{[{\bf k}]_2} \\
                               \vdots \\
                                \gamma_{[{\bf k}]_N}\\
                               \end{array}
                                \right)
                             \qquad  {\boldsymbol{\varepsilon}}_{{n_1,n_2,\cdots, n_N}}:=\left(
                              n_{1}, n_{2},  \cdots,
n_{N}
\right)
\eeq
such that
\beq{\label{quik}}
{\boldsymbol{\varepsilon}}_{{n_1,n_2,\cdots, n_N}} \cdot {\boldsymbol{\gamma}}_{[\bf{k}]}  = \sum_{l=1}^{N}n_l \gamma_{[{\bf k}]_l}.
\eeq

From the computation
\beq
\fl \langle J_{[{\bf k}]_l},\gamma_{[{\bf k}]_l}|J_{[{\bf k}]_l},\gamma_{[{\bf k}]_l}\rangle &=&
(\mathcal N(J_{[{\bf k}]_l}))^{-1}\sum_{n_{l},n_{q}=0}^{\infty}\frac{J^{(n_l + n_{q})/2}_{[{\bf
k}]_l}e^{-\imath (n_{l} - n_{q})\gamma_{[{\bf k}]_l}}}{\sqrt{n_l ! n_{q} !}} 
\langle \Psi_{[{\bf k}]}|\Psi_{[{\bf k}]}\rangle \otimes \langle \Phi^{[{\bf k}]}_{n_q}|\Phi^{[{\bf k}]}_{n_l}\rangle \crcr
&=& (\mathcal N(J_{[{\bf k}]_l}))^{-1}\sum_{n_{l}=0}^{\infty} \frac{J^{n_l}_{[{\bf k}]_l}}{n_l !}
\eeq
 we conclude that the coherent states
 $|J_{[{\bf k}]_l},\gamma_{[{\bf k}]_l}\rangle$ are orthonormalized
  if and only if the normalization constant $\mathcal N(J_{[{\bf k}]_l}) = e^{J_{[{\bf k}]_l}}.$

The multidimensional coherent states
(\cite{gazeau-novaes}) $|{\bf
J}_{[\bf{k}]},\gamma_{[\bf{k}]}\rangle$ are given by
\beq{\label{la}} 
\fl |{\bf
J}_{[\bf{k}]},\gamma_{[\bf{k}]}\rangle &=&
\left[\exp\left(\sum_{l=1}^{N}J_{[{\bf k}]_l}\right)\right]^{-1/2}
\prod_{l=1}^{N}\sum_{n_l=0}^{\infty} \frac{J^{n_l/2}_{[{\bf
k}]_l}e^{-\imath n_l \gamma_{[{\bf k}]_l}}}{\sqrt{n_l !}}
|\Psi_{[\bf k]}\rangle \otimes |\Phi^{[\bf k]}_{n_1,n_2,\cdots,
n_N}\rangle \eeq or, equivalently, using (\ref{silk}) and (\ref{quik}) and  setting
\beq \mathcal N({\bf J}_{[\bf k]}) =
\prod_{l=1}^{N} \mathcal N(J_{[{\bf k}]_l}) =
\exp\left(\sum_{l=1}^{N}J_{[{\bf k}]_l}\right) \eeq
  \beq{\label{i}} |{\bf J}_{[\bf{k}]},{\boldsymbol{\gamma}}_{[\bf{k}]}\rangle=
(\mathcal N({\bf J}_{[\bf k]}))^{-1/2}\sum_{[\bf n]=0}^{\infty}
 \frac{{\bf J}^{{\bf{n}}/2}_{[\bf k]}e^{-\imath {\boldsymbol{\varepsilon}}_{[{\bf
n}]}{\boldsymbol{\gamma}}_{[\bf k]}}}{\sqrt{\bf{n} !}}|\Psi_{[\bf k]}\rangle \otimes |\Phi^{[\bf k]}_{[\bf n]}\rangle
\eeq
where ${\boldsymbol{\varepsilon}}_{[{\bf
n}]} := {\boldsymbol{\varepsilon}}_{{n_1,n_2,\cdots, n_N}}, \,
 {\bf n} ! := n_1 !n_2 !\cdots n_N !$ and $ [{\bf n}] := n_1,n_2,\cdots, n_N.$

There follow the vector coherent states
\beq{\label{j}}
|\mathfrak J,{\boldsymbol{\gamma}};[\bf k]\rangle &=&(\mathcal N({\bf J_{[{\bf k}]}}))^{-1/2}\sum_{[\bf n]=0}^{\infty}\frac
{\mathfrak J^{{\bf n}/2}e^{-\imath {\boldsymbol{\varepsilon \gamma}}}}{\sqrt{\bf{n} !}}|\Xi_{[{\bf n}];[{\bf k}]}\rangle \cr
                  &=&   \left(
                      \begin{array}{c}
                         0 \\
                        \vdots \\
                         0 \\
                         |{\bf J}_{[\bf k]},\gamma_{[\bf k]}\rangle \\
                        0 \\
                        \vdots \\
                        0 \\
                       \end{array}
                        \right)
\eeq
where:

\beq{\label{diagonal}}
&& \mathfrak J = {\rm diag}({\bf J}_{(0,0, \dots, 0,0)}, {\bf J}_{(1,0, \dots 0,0)}, \dots , {\bf J}_{(1,1, \dots, 1,1)}) \crcr
&& {\boldsymbol{\gamma}} = {\rm diag}( {\boldsymbol{\gamma}}_{(0,0, \dots, 0,0)}, 
 {\boldsymbol{\gamma}}_{(1,0, \dots, 0,0)}, \dots , {\boldsymbol{\gamma}}_{(1,1, \dots, 1,1)}) \crcr
&& {\boldsymbol{\varepsilon}} = {\rm diag}({\boldsymbol{\varepsilon}}_{{[{\bf n}]}},{\boldsymbol{\varepsilon}}_{{[{\bf n}]}},\dots, 
{\boldsymbol{\varepsilon}}_{{[{\bf n}]}})
\eeq 
and
\beq
|\Xi_{[{\bf n}];[{\bf k}]}\rangle := \left(
                      \begin{array}{c}
                         0 \\
                        \vdots \\
                         0 \\
                         |\Psi_{[\bf k]} \otimes \Phi^{[\bf k]}_{[\bf n]}\rangle \\
                        0 \\
                        \vdots \\
                        0 \\
                       \end{array}
                        \right).
\eeq
The Hamiltonian $H_{D}$
splits into $2^{M}$ orthogonal parts $\tilde H_{[\bf k]}$.
Since the lowest eigenvalue $ E^{[{\bf k}]}_0 $ of the component
$\tilde H_{[\bf k]}$ is zero for $ k_i=0, i=1, 2,\cdots, M $, we consider
the Hamiltonian  $H'_{D}$ such that \beq{\label{le}} 
\fl H'_{D} =
\bigoplus_{[{\bf k}] \in \Gamma} \tilde H'_{[\bf k]} \qquad \tilde H'_{[\bf k]}
= \sum_{[\bf n]=0}^{\infty}\left(E^{[\bf k]}_{[\bf n]}-E^{[\bf k]}_{[\bf
0]}\right)|\Psi_{[\bf k]} \otimes \Phi^{[\bf k]}_{[\bf n]}\rangle \langle \Psi_{[\bf k]} \otimes \Phi^{[\bf k]}_{[\bf n]}| \eeq
where $\Gamma$ is as in (\ref{set00}).
In (\ref{d}), we define
$\Omega_{N}$,  the average frequency of the $N$ bosonic modes,  as
\beq{\label{omega}} \Omega_{N} :=  \left(\sum_{l=1}^{N}\omega_{l}
n_{l}\right)\left/ \left(\sum_{l=1}^{N} n_{l}\right)\right. \eeq transforming (\ref{d}) into
\beq{\label{lil}} E^{[{\bf k}]}_{[\bf n]} =
\Omega_{N}\left(\sum_{l=1}^{N} n_l + \frac{\epsilon_{[{\bf
k}]}}{\Omega_{N}} - \frac{g^2_{[{\bf
k}]}}{\Omega_{N}}\sum_{l=1}^{N}\frac{1}{\omega_l}\right). \eeq The
vector coherent states given in (\ref{j})
 satisfy the required properties, i.e.:

 $(i)$ the
temporal stability,
 \beq{\label{stab}}
e^{-\imath H'_{D} t}|\mathfrak J,{\boldsymbol{\gamma}}; [{\bf k}]\rangle = |\mathfrak J,{\boldsymbol{\gamma}} +
\Omega_{N}t \beta d_{[{\bf k}]}; {\bf{[k]}}\rangle \eeq
where $d_{[{\bf k}]}$ is the diagonal matrix with $1$ in the $[{\bf
k}][{\bf k}]$-position and $0$ elsewhere, and   $\beta$ standing for a column matrix with $N$ lines, given by
\beq{\label{mat00}}
\beta = \left(
                      \begin{array}{c}
                         1 \\
                        \vdots \\
                         1 \\
                        \vdots \\
                        1 \\
                       \end{array}
                        \right)
\eeq
$(ii)$ the
action identity,
\beq{\label{identy}} \langle
\mathfrak J,{\boldsymbol{\gamma}};{\bf{[k]}}|H'_{D}|\mathfrak J,{\boldsymbol{\gamma}};{\bf{[k]}}\rangle
=\Omega_{N} \sum_{l=1}^{N}J_{[{\bf k}]_{l}} \eeq
and
$(iii)$ the resolution of the identity
\beq{\label{resolution}}
\sum_{[{\bf k}]\in \Gamma}\int_{\mathcal D^{2^{M}}}|\mathfrak J,{\boldsymbol{\gamma}};[{\bf k}]\rangle 
\langle \mathfrak J,{\boldsymbol{\gamma}};[{\bf k}]| \mathcal N({\bf{J_{[{\bf k}]}}})d\mu({\boldsymbol{\gamma}})d\nu(\mathfrak J)
= \mathbb I_{2^{M}} \otimes I_{\hat\mathfrak H} \eeq
 where on each subspace of $\C^{2^{M}} \otimes \hat \mathfrak H$,
\beq
\int_{\mathcal D}|{\bf J}_{[\bf k]},{\boldsymbol{\gamma}}_{[\bf k]}\rangle \langle {\bf J}_{[\bf k]},{\boldsymbol{\gamma}}_{[\bf k]}|
 \mathcal N({\bf{J_{[{\bf k}]}}})d\mu({\boldsymbol{\gamma}}_{[\bf k]})d\nu_{[{\bf k}]}({\bf J}_{[\bf k]}) = \mathbb P_{[{\bf k}]}
\eeq
holds with the measures given by $d\mu({\boldsymbol{\gamma}}) = \prod_{[{\bf k}] \in \Gamma}d\mu({\boldsymbol{\gamma}}_{[\bf k]})$, 
$d\nu(\mathfrak J) = \prod_{[{\bf k}] \in \Gamma}d\nu_{[{\bf k}]}({\bf J}_{[\bf k]})$
and $\mathcal D = \prod_{k=1}^{N}([0,\infty) \times [0,2\pi))^{k}$;  the measures 
$d\nu_{l}(J_{[{\bf k}]_{l}}) = e^{-J_{[{\bf k}]_{l}}}dJ_{[{\bf k}]_{l}}, 1\leq l\leq N$ solve the following moment problems:
\beq
\frac{1}{n_{l} !}\int_{0}^{\infty}J^{n_{l}}_{[{\bf k}]_{l}}d\nu_{l}(J_{[{\bf k}]_{l}}) = 1 \qquad \int_{0}^{\infty}d\nu_{l}(J_{[{\bf k}]_{l}}) = 1.
\eeq

\subsection{Vector coherent states for the nondegenerate Hamiltonian with a general  spectrum}
We discuss now the vector coherent states  construction for the  Hamiltonian assumed to have  a  general positive energy 
spectrum 
provided by  $E^{[{\bf k}]}_{lj} - E^{[{\bf k}]}_{l0} =
\omega_{l} \epsilon_{lj},  
\; 1\leq l \leq N, \; j=0, 1, 2, \dots, \infty$ where the conditions  $\epsilon_{l0} = 0$  and 
$\epsilon_{lj} = \epsilon_{l'j'}$ iff $ l = l'$
and $j = j'$ are satisfied. Let $\mathfrak H_{[{\bf k}]} = span\{
\Phi^{[{\bf k}]}_{[Nj]}  \otimes \Psi_{[{\bf k}]},  
\quad  |\Phi^{[{\bf k}]}_{[Nj]} \rangle = 
\bigotimes^{N}_{l=1}|\Phi^{[{\bf k}]}_{lj} \rangle, j \in  \N \}$ where $[Nj]:=  1j, 2j, \cdots, Nj$.   
Furthermore, consider that all the $\omega_l$ are different for all $
l, 1\leq l \leq N$ with the information about the degeneracy obtained as previously. 

Setting \beq
  \epsilon_{lj} = \frac{E^{[{\bf k}]}_{lj} - E^{[{\bf k}]}_{l0}}{\omega_l} \eeq
we consider  a general  sequence of dimensionless quantities   given by $
(\epsilon_{lj})_{j \in \N}$ satisfying the inequalities $
0=\epsilon_{l0} <\epsilon_{l1} <\epsilon_{l2} <\dots.$

Taking $l, 1\leq l \leq
N,$ we define the  coherent states  $ |J_{[{\bf
k}]_l},\gamma_{[{\bf k}]_l}\rangle$,  where $ 0 \leq  J_{[{\bf k}]_l} < L_{l} 
= \lim_{j \rightarrow \infty} \epsilon_{lj}$ 
and $\gamma_{[{\bf k}]_l} \in \R$, as follows: \beq
\fl |J_{[{\bf k}]_l},\gamma_{[{\bf k}]_l}\rangle = (\mathcal N(J_{[{\bf k}]_l}))^{-1/2}\sum_{j=0}^{\infty}
\frac{J^{{j}/2}_{[{\bf
k}]_l}e^{-\imath \epsilon_{lj} \gamma_{[{\bf k}]_l}  }}{\sqrt{\rho_{l}(j)}}
|\Psi_{[{\bf k}]}\rangle \otimes |\Phi^{[{\bf k}]}_{lj} \rangle \qquad \rho_{l}(j) = \prod^{j}_{q=1}\epsilon_{lq}
\eeq 
with the condition $\rho_{l}(0) = 1$.  The normalization factor $\mathcal N(J_{[{\bf k}]_l})$ is chosen such
that the set of states $\{|J_{[{\bf k}]_l},\gamma_{[{\bf k}]_l}\rangle \}_{l=1}^{N}$ forms an orthonormal basis, i.e.
\beq
\langle J_{[{\bf k}]_l},\gamma_{[{\bf k}]_l}|J_{[{\bf k}]_m},\gamma_{[{\bf k}]_m}\rangle = \delta_{lm} \qquad \qquad
\mathcal N(J_{[{\bf k}]_l}) = \sum_{j=0}^{\infty}
\frac{J^{j}_{[{\bf
k}]_l}}{\rho_{l}(j)}.
\eeq

Let
\beq
{\bf{J}}_{[\bf{k}]} := \prod_{l=1}^{N} J_{[{\bf k}]_l} \qquad  \qquad
{\boldsymbol{\varepsilon}}_{[Nj]}:=\left(
                              \epsilon_{1j}, \epsilon_{2j},  \cdots,
\epsilon_{Nj}
\right)
\eeq
 such that, with ${\boldsymbol{\gamma}}_{[\bf{k}]} $ given as in (\ref{silk}),  we get 
\beq
 {\boldsymbol{\varepsilon}}_{[Nj]} \cdot {\boldsymbol{\gamma}}_{[\bf{k}]}  
= \sum_{l=1}^{N}\epsilon_{lj} \gamma_{[{\bf k}]_l}.
\eeq

The multidimensional coherent states $|{\bf
J}_{[\bf{k}]},\gamma_{[\bf{k}]}\rangle$ are given by
\beq 
\fl |{\bf
J}_{[\bf{k}]},\gamma_{[\bf{k}]}\rangle &=&
\left[\prod_{l=1}^{N} \mathcal N(J_{[{\bf k}]_l})\right]^{-1/2}
\prod_{l=1}^{N}\sum_{j=0}^{\infty} \frac{J^{j/2}_{[{\bf
k}]_l}e^{-\imath \epsilon_{lj} \gamma_{[{\bf k}]_l}}}{\sqrt{\rho_{l}(j)}}
|\Psi_{[{\bf k}]}\rangle \otimes |\Phi^{[{\bf k}]}_{[Nj]} \rangle \eeq 
or, 
equivalently 

 \beq 
\fl |{\bf J}_{[\bf{k}]},{\boldsymbol{\gamma}}_{[\bf{k}]}\rangle=
(\mathcal N({\bf J}_{[\bf k]}))^{-1/2}\sum_{j=0}^{\infty} 
 \frac{{\bf J}^{{j}/2}_{[\bf k]}
e^{-\imath {\boldsymbol{\varepsilon}}_{[Nj]}.{\boldsymbol{\gamma}}_{[\bf k]}}}{\sqrt{\rho(j)}}
|\Psi_{[\bf k]}\rangle \otimes |\Phi^{[{\bf k}]}_{[Nj]} \rangle  \qquad  \rho(j) = \prod^{N}_{l=1}\rho_{l}(j).
\eeq

The vector coherent states are now given as follows:
\beq{\label{vectcs}}
|\mathfrak J,{\boldsymbol{\gamma}};[\bf k]\rangle &=&(\mathcal N({\bf J_{[{\bf k}]}}))^{-1/2}\sum_{j=0}^{\infty}
  \frac
{\mathfrak J^{j/2}e^{-\imath {\boldsymbol{\varepsilon \gamma}}}}{\sqrt{\rho(j)}}|\Xi_{[Nj];[{\bf k}]}\rangle \cr
                  &=&   \left(
                      \begin{array}{c}
                         0 \\
                        \vdots \\
                         0 \\
                         |{\bf J}_{[\bf k]},\gamma_{[\bf k]}\rangle \\
                        0 \\
                        \vdots \\
                        0 \\
                       \end{array}
                        \right)
\eeq
where $ \mathfrak J $ and   ${\boldsymbol{\gamma}}$ are  provided  as in (\ref{diagonal}), 

\beq
{\boldsymbol{\varepsilon}} = {\rm diag}({\boldsymbol{\varepsilon}}_{{[Nj]}},{\boldsymbol{\varepsilon}}_{{[Nj]}},\dots, 
{\boldsymbol{\varepsilon}}_{{[Nj]}})
\eeq 
and
\beq
|\Xi_{[Nj];[{\bf k}]}\rangle := \left(
                      \begin{array}{c}
                         0 \\
                        \vdots \\
                         0 \\
                         |\Psi_{[\bf k]} \otimes \Phi^{[\bf k]}_{[Nj]}\rangle \\
                        0 \\
                        \vdots \\
                        0 \\
                       \end{array}
                        \right).
\eeq
Assume that 
the Hamiltonian  $H'_{D}$ admits the following spectral decomposition  \beq
\fl H'_{D} =
\bigoplus_{[{\bf k}] \in \Gamma} \tilde H'_{[\bf k]} \qquad \tilde H'_{[\bf k]}
= \sum_{j=0}^{\infty}\left(E^{[\bf k]}_{[Nj]}-E^{[\bf k]}_{[
N0]}\right)|\Psi_{[\bf k]} \otimes \Phi^{[\bf k]}_{[Nj]}\rangle \langle \Psi_{[\bf k]} \otimes \Phi^{[\bf k]}_{[Nj]}| \eeq
where $E^{[\bf k]}_{[Nj]}-E^{[\bf k]}_{[
N0]} = \Omega \sum_{l=1}^{N}\epsilon_{lj}$, $\Omega$ denoting the average frequency of the $N$ bosonic modes.

The
vector coherent states given in (\ref{vectcs})
 satisfy the required properties, i.e.:

 $(i)$ the
temporal stability,
 \beq{\label{genstab}}
e^{-\imath H'_{D} t}|\mathfrak J,{\boldsymbol{\gamma}}; [{\bf k}]\rangle = |\mathfrak J,{\boldsymbol{\gamma}} +
\Omega t \beta d_{[{\bf k}]}; {\bf{[k]}}\rangle \eeq
where $d_{[{\bf k}]}$ and   $\beta$ are  given as before

$(ii)$ the
action identity,
\beq{\label{genident}} 
\langle
\mathfrak J,{\boldsymbol{\gamma}};{\bf{[k]}}|H'_{D}|\mathfrak J,{\boldsymbol{\gamma}};{\bf{[k]}}\rangle
=\Omega \sum_{l=1}^{N}J_{[{\bf k}]_{l}} \eeq
and
$(iii)$ the resolution of the identity
\beq{\label{genres00}} 
\sum_{[{\bf k}]\in \Gamma}\int_{\mathcal D^{2^{M}}}|\mathfrak J,{\boldsymbol{\gamma}};[{\bf k}]\rangle 
\langle \mathfrak J,{\boldsymbol{\gamma}};[{\bf k}]| \mathcal N({\bf{J_{[{\bf k}]}}})d\mu_{B}({\boldsymbol{\gamma}})d\nu(\mathfrak J)
= \mathbb I_{2^{M}} \otimes I_{\hat\mathfrak H} \eeq
$\hat \mathfrak H$ being  spanned
by the bosonic states $\left\{|\Phi^{[{\bf k}]}_{[Nj]}\rangle,  [{\bf k}] \in \Gamma \right\}_{j=0}^{\infty}$,  
where on each subspace of $\C^{2^{M}} \otimes \hat \mathfrak H$,
\beq{\label{genres01}} 
\fl&&\int_{0}^{L_1}\int_{0}^{L_2}
\cdots \int_{0}^{L_N} \left[\int_{\R}\int_{\R}\cdots \int_{\R}
|{\bf J}_{[\bf k]},{\boldsymbol{\gamma}}_{[\bf k]}\rangle \langle {\bf J}_{[\bf k]},{\boldsymbol{\gamma}}_{[\bf k]}|
 \mathcal N({\bf{J_{[{\bf k}]}}})d\mu_{B}({\boldsymbol{\gamma}}_{[\bf k]})\right]d\nu_{[{\bf k}]}({\bf J}_{[\bf k]}) \cr 
\fl&&= \sum_{j=0}^{\infty}
| \Psi_{[{\bf k}]} \otimes \Phi^{[{\bf k}]}_{[Nj]}\rangle \langle \Psi_{[{\bf k}]} \otimes \Phi^{[{\bf k}]}_{[Nj]} | 
\eeq
holds with  $d\mu_{B}({\boldsymbol{\gamma}}) = \prod_{[{\bf k}] \in \Gamma}d\mu_{B}({\boldsymbol{\gamma}}_{[\bf k]})$, $d\nu(\mathfrak J)$ given  
as 
before 
and $\mathcal D = \prod_{l=1}^{N} [0, L_{l}) \times \R^{N}$;  the related measures 
$d\nu_{l}(J_{[{\bf k}]_{l}}) $ are assumed to 
solve the following moment problems:
\beq
\frac{1}{\rho_{l}(j)}\int_{0}^{L_{l}}J^{j}_{[{\bf k}]_{l}}d\nu_{l}(J_{[{\bf k}]_{l}}) = 1 \qquad 
\int_{0}^{L_{l}}d\nu_{l}(J_{[{\bf k}]_{l}}) = 1.
\eeq

The measures $d\mu_{B}(\gamma_{[{\bf k}]_{l}}),  1\leq l \leq
N$,  refer to the {\it Bohr measure} $d\mu_{B}$  \cite{ab}.


\subsection{Vector coherent states for the degenerate Hamiltonian}

Here, we extend the preceding construction to the situation in which the eigenvalues
 of the  Hamiltonian  $H$ with discrete positive spectrum are degenerate, the degeneracies being finite.
Following \cite{ab}, in order to meet the Gazeau-Klauder formalism,
 we introduce a third parameter into the definition of the coherent states,
replacing  $|J,\gamma\rangle$  by $|J,\gamma,\theta\rangle$ and $|J_{[{\bf k}]},\gamma_{[{\bf k}]}\rangle$
by $|J_{[{\bf k}]},\gamma_{[{\bf k}]},\theta\rangle.$

For $\omega_l = \omega,$ with $l=1, 2, \dots, N$, the eigenvalues
$ E^{[{\bf k}]}_{n_1,n_2,\dots, n_N}$
are degenerate.
Set $d(n) = C^{n}_{n+N-1} = \frac{(n+N-1) !}{n !(N-1) !}$,
the degree of degeneracy of the $n$th energy level.  The equations  (\ref{d})  and (\ref{e}) with
$n= n_1+ n_2 +\dots +n_N $ become
\beq
E^{[{\bf k}]}_n  &=& \omega n + \epsilon_{[{\bf k}]}- N\frac{g^2_{[{\bf k}]}}{\omega} \\
\varphi^{[{\bf k}]}_n \equiv \varphi^{[{\bf k}]}_{n,j} &=&
\Phi^{[{\bf k}]}_{n,j}\otimes \Psi_{[{\bf k}]} \qquad j=1,2, \cdots,
d(n). \eeq We have  $E^{[{\bf k}]}_n - E^{[{\bf k}]}_{0} = \omega
n$. Replacing  $n$ by $\epsilon_n, \quad\forall n \geq 0$, it comes
\beq{\label{su}} \epsilon_{n} = \frac{E^{[{\bf k}]}_n-E^{[{\bf
k}]}_{0}}{\omega} \eeq which leads to  a sequence of dimensionless
quantities  $(\epsilon_n)_{n \in \N}$ such that  
$0 = \epsilon_0 <
\epsilon_1<\epsilon_2<\epsilon_3<\dots $. 
Then,  by analogy with the previous result (see (\ref{le})) with the same arguments, the Hamiltonian $H$ can be reduced 
to the corresponding counterpart
  $\hat H'_{D}$ as follows:
\beq
\fl \hat H'_{D} = \bigoplus_{[{\bf k}]\in \Gamma}\hat H'_{[{\bf k}]} \qquad  \hat H'_{[{\bf k}]}=
\omega\sum_{n=0}^{\infty}\sum_{j=1}^{d(n)}\epsilon_{n}|\Psi_{[{\bf k}]} \otimes
\Phi^{[{\bf k}]}_{n,j}\rangle\langle\Psi_{[{\bf k}]}\otimes
\Phi^{[{\bf k}]}_{n,j}| \qquad \hbar = 1
\eeq
 generating, for the parameter $\theta \in [0,2\pi)$,  the coherent
states  defined  by \cite{ab}
\beq{\label{vcs00}} |J_{[{\bf k}]}, \gamma_{[{\bf k}]}, \theta\rangle =
(\mathcal N(J_{[{\bf k}]}))^{-1/2}\sum_{n=0}^{\infty}\sum_{j=1}^{d(n)}
\frac{J^{n/2}_{[{\bf k}]}e^{-\imath n \gamma_{[{\bf k}]}}e^{-\imath
j \theta}}{\sqrt{n ! d(n)}}|\Psi_{[{\bf k}]}\rangle\otimes
|\Phi^{[{\bf k}]}_{n,j}\rangle  \eeq  such that
$\langle J_{[{\bf k}]}, \gamma_{[{\bf k}]}, \theta|J_{[{\bf k}]}, \gamma_{[{\bf k}]}, \theta\rangle = 1$ for

\beq
\mathcal N(J_{[{\bf k}]}) =  e^{J_{[{\bf k}]}}.
\eeq
In this  representation, the vectors  $\Psi_{[{\bf
k}]}$ form the canonical basis of $\C^{2^M}$ as noticed in
(\ref{base}).

Setting  $|\Phi^{[{\bf k}]}_{n,j}\rangle \equiv |\Phi^{[{\bf k}]}_{j-1,d(n)-j}\rangle,\,j=1, 2, \dots, d(n)$,
the coherent states can be rewritten as
\beq{\label{f}}
|J_{[{\bf k}]}, \gamma_{[{\bf k}]}, \theta\rangle = e^{-\frac{J_{[{\bf k}]}}{2}}\sum_{n=0}^{\infty}\sum_{j=1}^{d(n)}
\frac{J^{n/2}_{[{\bf k}]}e^{-\imath n \gamma_{[{\bf k}]}}e^{-\imath j \theta}}{\sqrt{n ! d(n)}}
|\Psi_{[{\bf k}]}\rangle \otimes |\Phi^{[{\bf k}]}_{j-1,d(n)-j}\rangle.
\eeq
At this stage of our development, let us make clearer the meaning of the bosonic states
 $|\Phi^{[{\bf k}]}_{j-1,d(n)-j}\rangle$ considering some values of the number of vibrational modes $N$ as below:
\bitem
\item [$(i)$]  
$N=2$, for all $n\geq 1$,
\beq{\label{exemlp00}}
\fl |\Phi^{[{\bf k}]}_{j-1,d(n)-j}\rangle =
\frac{ (A^{\dag}_{[{\bf k}]_1})^{j-1}e^{\imath \sqrt{2}\frac{g_{[{\bf k}]}}{\omega} P_1} \otimes (A^{\dag}_{[{\bf k}]_2})^{d(n)-j}
e^{\imath \sqrt{2}\frac{g_{[{\bf k}]}}{\omega} P_2}}{\sqrt{(j-1)! (d(n)-j)!}} 
(|0\rangle \otimes |0\rangle).
\eeq
\item [$(ii)$] $N=3$, for all $n\geq 1$, the  bosonic states  
$|\Phi^{[{\bf k}]}_{j-1,d(n)-j,0}\rangle$, 
$|\Phi^{[{\bf k}]}_{0,j-1, d(n)-j}\rangle$ and $|\Phi^{[{\bf k}]}_{j-1,0,d(n)-j}\rangle$  
are 
respectively 
given by
\beq{\label{exempl01}} 
&& \fl |\Phi^{[{\bf
k}]}_{j-1,d(n)-j,0}\rangle =\frac{(A^{\dag}_{[{\bf k}]_1})^{j-1}e^{\imath
\sqrt{2}\frac{g_{[{\bf k}]}}{\omega} P_1} \otimes (A^{\dag}_{[{\bf
k}]_2})^{d(n)-j}e^{\imath \sqrt{2}\frac{g_{[{\bf k}]}}{\omega} P_2}
\otimes e^{\imath \sqrt{2}\frac{g_{[{\bf k}]}}{\omega} P_3} }{\sqrt{(j-1)! (d(n)-j)!}} \crcr 
&& \times (|0\rangle \otimes |0\rangle \otimes |0\rangle),
\eeq 
 
\beq{\label{exempl02}} 
&& \fl |\Phi^{[{\bf
k}]}_{0,j-1,d(n)-j}\rangle = \frac{e^{\imath \sqrt{2}\frac{g_{[{\bf k}]}}{\omega} P_1} \otimes (A^{\dag}_{[{\bf k}]_2})^{j-1} e^{\imath
\sqrt{2}\frac{g_{[{\bf k}]}}{\omega} P_2}\otimes (A^{\dag}_{[{\bf
k}]_3})^{d(n)-j}e^{\imath \sqrt{2}\frac{g_{[{\bf k}]}}{\omega}
P_3}}{\sqrt{(j-1)! (d(n)-j)!}} \crcr 
&& \times (|0\rangle \otimes |0\rangle \otimes |0\rangle), \eeq
 
\beq{\label{exempl03}} 
\fl |\Phi^{[{\bf k}]}_{j-1,0, d(n)-j}\rangle &=& \frac{
(A^{\dag}_{[{\bf
k}]_1})^{j-1}e^{\imath \sqrt{2}\frac{g_{[{\bf k}]}}{\omega} P_1} \otimes e^{\imath \sqrt{2}\frac{g_{[{\bf k}]}}{\omega} P_2} \otimes 
(A^{\dag}_{[{\bf k}]_3})^{d(n)-j}e^{\imath \sqrt{2}\frac{g_{[{\bf
k}]}}{\omega} P_3}
}{\sqrt{(j-1)! (d(n)-j)!}} \crcr
&& \times (|0\rangle \otimes |0\rangle \otimes |0\rangle). \eeq 
\eitem
 In the sequel, for
the notation simplification,  $|\Phi^{[{\bf k}]}_{j-1,d(n)-j}\rangle$
will denote a bosonic state for all values of $N.$

The vector coherent states related to the coherent states (\ref{f}) are defined  on
the Hilbert space  $\C^{2^{M}} \otimes \tilde{{\mathfrak H}}, $ where $\tilde{\mathfrak H}$
is spanned by the bosonic states  $|\Phi^{[{\bf k}]}_{j-1,d(n)-j}\rangle$ such that
\beq{\label{gi}}
|\mathfrak J, {\boldsymbol{\gamma}}, \theta; [{\bf k}]\rangle = e^{-\frac{ J_{[{\bf k}]}}{2}}\sum_{n=0}^{\infty}\sum_{j=1}^{d(n)}
\frac{\mathfrak J^{n/2}e^{-\imath n {\boldsymbol{\gamma}}}e^{-\imath j \theta}}{\sqrt{n ! d(n)}}
|\chi_{j-1,d(n)-j;[{\bf k}]}\rangle
\eeq
with:
\beq{\label{diago}}
\mathfrak J &=& {\rm diag}(J_{(0,0,\dots, 0,0)}, J_{(1,0, \dots, 0,0)}, \dots, J_{(1,1,\dots, 1,1)}) \crcr
{\boldsymbol{\gamma}}&=& {\rm diag}({\gamma}_{(0,0,\dots, 0,0)}, {\gamma}_{(1,0, \dots, 0,0)}, \dots, 
 {\gamma}_{(1,1,\dots, 1,1)})
\eeq

and

\beq
|\chi_{j-1,d(n)-j;[{\bf k}]}\rangle := \left(
                      \begin{array}{c}
                         0 \\
                        \vdots \\
                         0 \\
                         |\Psi_{[{\bf k}]} \otimes \Phi^{[{\bf k}]}_{j-1,d(n)-j}\rangle\\
                        0 \\
                        \vdots \\
                        0 \\
                       \end{array}
                        \right).
\eeq

The vector coherent states (\ref{gi})
also fulfill the temporal stability, action identity and  resolution of the identity properties, with the 
measures given by  
$d\nu_{N}(J_{[{\bf k}]}) = \left[\frac{e^{-J_{[{\bf k}]}}J^{N-1}_{[{\bf k}]}}{(N-1)!} + J^{N}_{[{\bf k}]} \delta (J_{[{\bf k}]}) \right]
dJ_{[{\bf k}]}$ associated with the  moment problems
\beq
\frac{1}{n !d(n)}\int_{0}^{\infty}J_{[{\bf k}]}^{n}d\nu_{N}(J_{[{\bf k}]}) = 1 \qquad n=0,1,2, \dots.
\eeq

\subsection{Vector coherent states for the degenerate Hamiltonian with a general spectrum}
Assume that the degenerate Hamiltonian  has a general  positive energy spectrum 
provided by  $E^{[{\bf k}]}_{n} - E^{[{\bf k}]}_{0} =
\omega \epsilon_{n}$  with a finite degree of degeneracy $d(n)$ for all $n$. Consider the general sequence $0 = \epsilon_{0} < 
\epsilon_{1}  <  \epsilon_{2} < \epsilon_{3} < \dots $. According to Sections $2.2$ and $2.3$, the coherent states, where $0 \leq 
J_{[{\bf k}]} < L, \gamma_{[{\bf k}]} \in \R$ and  $\theta \in [0, 2\pi)$,  are given by
\beq{\label{vcsdeg}}
\fl |J_{[{\bf k}]}, \gamma_{[{\bf k}]}, \theta\rangle = 
(\mathcal N(J_{[{\bf k}]}))^{-1/2}\sum_{n=0}^{\infty}\sum_{j=1}^{d(n)}
\frac{J^{n/2}_{[{\bf k}]}e^{-\imath \epsilon_{n} \gamma_{[{\bf k}]}}e^{-\imath j \theta}}{\sqrt{\epsilon_{n}!d(n)}}
|\Psi_{[{\bf k}]}\rangle \otimes |\Phi^{[{\bf k}]}_{j-1,d(n)-j}\rangle
\eeq
where $\epsilon_{n}! := \epsilon_{1}\epsilon_{2} \cdots \epsilon_{n}$.

The normalization factor by use of the relation 
$\langle J_{[{\bf k}]}, \gamma_{[{\bf k}]}, \theta|J_{[{\bf k}]}, \gamma_{[{\bf k}]}, \theta   \rangle = 1$ is provided as 
\beq
\mathcal N(J_{[{\bf k}]}) = \sum_{n=0}^{\infty} \frac{J^{n}_{[{\bf k}]}}{\epsilon_{n}!}
\eeq
which is a power series in $J$ with $L > 0$ assumed to be its radius of convergence.

Then, the corresponding  vector coherent states  are   given as follows:

\beq{\label{VCSdeg}}
\fl |\mathfrak J, {\boldsymbol{\gamma}}, \theta; [{\bf k}]\rangle = (\mathcal N(J_{[{\bf k}]}))^{-1/2}
\sum_{n=0}^{\infty}\sum_{j=1}^{d(n)}
\frac{\mathfrak J^{n/2}e^{-\imath \epsilon_{n} {\boldsymbol{\gamma}}}e^{-\imath j \theta}}{\sqrt{\epsilon_{n} ! d(n)}}
|\chi_{j-1,d(n)-j;[{\bf k}]}\rangle
\eeq

where $\mathfrak J, {\boldsymbol{\gamma}}$ are given by (\ref{diago}). 

The vector coherent states (\ref{VCSdeg})
 fulfill the temporal stability, action identity and  resolution of the identity properties, with the 
measures  $d\nu(J_{[{\bf k}]}) $ assumed to solve the  moment problems
\beq
\frac{1}{\epsilon_{n} !d(n)}\int_{0}^{L}J_{[{\bf k}]}^{n}d\nu(J_{[{\bf k}]}) = 1.
\eeq

\section{Generalization}
This section addresses a method of building  coherent states
for a Hamiltonian model describing  a non radiative relaxation
mechanism caused by the inertial coupling of an electron to the
nanoparticle's translational center-of-mass (c.m.). This interaction is
present because an electron bound to an impurity  center in an
oscillating nanoparticle is in an accelerating reference frame, and,
in accordance with the Einstein's equivalence principle, it feels a
fictitious time-dependent force. Such a relaxation mechanism is
operative even at zero temperature, owing to the fact that quantum
zero-point motion of the c.m. is sufficient to produce the
fictitious force. See \cite{yang-geller-dennis} for more details.
This study is done in the context of
  low-energy  decay of a  localized electronic impurity state in a  macroscopic
semiconductor
  or insulator in condensed matter physics. The dissipation is followed by a phonon
emission.
The  model consists of a single nanoparticle of mass $m_{n}$ connected
to a bulk substrate by a few atomic bounds. The effect of the
substrate is to subject the nanoparticle to a one-dimensional
harmonic oscillator potential $V= \frac{1}{2}m_{n}\Omega^{2}X^{2}$ with
frequency $\Omega, $ the $X$ direction being perpendicular to the
plane of the substrate. The center of mass motion is that of a
macroscopic harmonic
 oscillator interacting with many other degrees of freedom, such as the phonons of
  the bulk substrate characterized by the creation and annihilation operators
$b^{\dag}, b.$ We restrict our study to the interaction between the center of mass of the nanoparticle and the fermion levels in the 
presence of the vibrational modes.
After a series of gauge transformations, the
general  Hamiltonian describing such a system, where
$\hbar = 1$, can be reduced to:  \beq\label{haml1} H_{CM} = &&
\sum_{n=1}^{N}\omega_{n}a^{\dag}_{n}a_{n} + \Omega b^{\dag}b
+\sum_{\alpha=1}^{M}\epsilon_{\alpha}c^{\dag}_{\alpha}c_{\alpha} -
g'\sum_{\alpha,\alpha'=1}^{M}x_{\alpha\alpha'}c^{\dag}_{\alpha}c_{\alpha'}(b+b^{\dag}).
\eeq
The first
term describes the nanoparticle's internal vibrational dynamics: the
$ \omega_{n}$ are the frequencies of the internal modes, and
$a^{\dag}_{n}$ and $a_{n}$ are the corresponding phonon creation and
annihilation operators. These internal vibrational modes have been
observed by low-frequency Raman scattering \cite{duval}-\cite{krauss} and
by femptosecond pump-probe spectroscopy \cite{cerullo}.
 The second term describes the harmonic dynamics of the center of
mass (c.m) of the nanoparticle which is assumed to be
constrained to move in the $x$ direction only. Hence, the c.m.
translational  motion is described by a single boson
degree-of-freedom \beq b = \sqrt{\frac{m_{n}\Omega}{2}} \left(X +
\frac{\imath}{m_{n}\Omega}P\right)\nonumber \eeq where $X$ and $P$ are
the $x$ components of the c.m. position and momentum.  The third term is
the Hamiltonian for a noninteracting $M$-level system.
The fourth term describes the inertial coupling between the $M$-level
system and the center of mass motion; here $x_{\alpha\alpha'} \equiv
\langle \phi_{\alpha}|x|\phi_{\alpha'} \rangle$ are dipole-moment
matrix elements, which, of course, depend on the form of the
impurity states.

In addition to the commutation rules given in (\ref{commuta}), one has
\beq{\label{regle}}
&&\fl [b,b^{\dag}] = \1 \qquad [b,b] = 0 \qquad [b^{\dag}, b^{\dag}]=0 \crcr
&& \fl [a^{\dag}_{i}, b] = 0 \qquad [a_{i},b^{\dag}]=0 \qquad
  [a^{\dag}_{i}, b^{\dag}] = 0  \qquad [a_{i},b ]=0 \qquad 1 \leq i  \leq N.
\eeq Before all things,  let us immediately clarify the terminology
problems to avoid any confusion. By {\it diagonal} (resp. {\it
extra-diagonal})  Hamiltonian, we always mean the Hamiltonian
encompassing all free boson and fermion contributions with only
diagonal elements in the interaction coupling terms (resp. with only
extra-diagonal elements in the interaction coupling terms).

\subsection{Diagonal case}

Given the commutation relations  (\ref{regle}),
setting $x_{\alpha\alpha'}  = \delta_{\alpha\alpha'}$ for
$\alpha, \alpha' = 1, 2, \dots, M$, the Hamiltonian (\ref{haml1})
can be transformed into the following form:
 \beq \label{Hdiag} H = H_{1} \otimes I_{\mathfrak H_{2}} + I_{\mathfrak H_{1}} \otimes H_{2}
\eeq
where
\beq
H_{1} = \sum_{i=1}^{N}\omega_{i}a^{\dag}_{i}a_{i}
\eeq
is the Hamiltonian of the internal
vibrational dynamics and
\beq
H_{2} = \Omega b^{\dag}b +
\sum_{\alpha=1}^{M}\epsilon_{\alpha}c^{\dag}_{\alpha} c_{\alpha} - g'\sum_{\alpha=1}^{M}
c^{\dag}_{\alpha}c_{\alpha}(b^{\dag} + b) \eeq
is the Hamiltonian of the c. m.  interacting with the $M$-levels;  $\mathfrak H_{1}$ and $\mathfrak H_{2}$ are 
the Hilbert spaces spanned by the eigenvectors of $H_{1}$ and $H_{2}$, respectively. The Hamiltonians $H_{1}$ and $H_{2}$ naturally commute, 
i.e.  $[H_{1}, H_{2}] = 0$, 
what can be easily checked using
 (\ref{commuta}) and (\ref{regle}).

Let us denote by
$\varphi = \chi \otimes \, \Psi_{[{\bf k}]}$ the eigenvector of $H_{2}$, where
 $[{\bf k}]= k_1,k_2,\dots, k_M$
with $k_1, k_2, \dots, k_M = 0, 1.$ So,
for all $M \geq 2 ,$ if there is not any zero in $[{\bf k}]$, then
\beq{\label{eqs}}
H_{2}\varphi=\left[ \sum_{j=1}^{M}
\delta_{k_{j},1}\epsilon_{j}  + \Omega
b^{\dag}b-Mg'(b+b^{\dag})\right]\varphi. \eeq In the contrary, if there exists
at least one zero in  $[{\bf k}]$, then the term $Mg'$ in (\ref{eqs}) is replaced by $\frac{g'}{2}\sum_{j=1}^{M}(1-\delta_{k_{j},0}+\delta_{k_{j},1})$.

As a matter of shortness of mathematical expressions, let us
also define the quantity $\kappa_{[{\bf k}]}$ such that \beq
\fl \kappa_{[{\bf k}]} = \cases{
              \begin{array}{lll}
              M \qquad \mbox{if} \,[{\bf k}]  \mbox{doesn't contain any zero}, \\
               \\
              \frac{1}{2} \sum_{j=1}^{M}(1-\delta_{k_{j},0}+\delta_{k_{j},1}) \qquad
\mbox{if}\, [{\bf k}]\,
\mbox{contains at least one zero.}
               \end{array}}
\eeq
By setting $g'\kappa_{[{\bf k}]}=g'_{\kappa_{[{\bf k}]}} $
 for all $[{\bf k}]$, we readily  find, as done  in the previous section, that
\beq{\label{op000}}
H_{2}\varphi  = \left[\Omega\left(b-\frac{g'_{\kappa_{[{\bf
k}]}}}{\Omega}\right)^{\dag}\left(b- \frac{g'_{\kappa_{[{\bf
k}]}}}{\Omega}\right)-\frac{{g'}^{2}_{\kappa_{[{\bf
k}]}}}{\Omega} + \epsilon_{[{\bf k}]}\right]\varphi.
\eeq

The eigenvectors given with the operator  \beq \mathcal C^{\dag} = e^{-\imath \sqrt{2}\frac{g'_{\kappa_{[{\bf
k}]}}}{\Omega}p}b^{\dag}e^{\imath \sqrt{2}\frac{g'_{\kappa_{[{\bf
k}]}}}{\Omega}p}=b^{\dag}-\frac{g'_{\kappa_{[{\bf
k}]}}}{\Omega}\eeq satisfying $\quad[b,\sqrt{2}p]=\imath$,  
are expressed as follows:
\beq
|\chi^{[\bf k]}_{m}\rangle = \frac{(\mathcal C^{\dag})^{m}}{\sqrt{m
!}}e^{-\imath \sqrt{2}\frac{g'_{\kappa_{[{\bf
k}]}}}{\Omega}p}|0\rangle \qquad m=0,1,2,\dots.\eeq

\subsubsection{Coherent states in the nondegenerate case.}
The Hamiltonian $H$ (\ref{Hdiag}) acts on a Hilbert
space spanned by the  tensor product of  the vectors $|\Phi_{[\bf n]}\rangle$ and $|\chi^{[\bf k]}_{m}\rangle \otimes |\Psi_{[\bf k]}\rangle$ which are
eigenvectors of the Hamiltonians  $H_{1}$
 and $H_{2},$ respectively. The coherent states
associated with these Hamiltonians are defined  on the Hilbert space $\mathfrak H_{1} \otimes \mathfrak H_{2}$. The eigenvalues of $H$  are given 
by \beq E^{[{\bf
k}]}_{[{\bf{n}}],m} = \sum_{l=1}^{N}\omega_l n_l  +  \left(\Omega m -
\frac{g'^{2}_{\kappa_{[{\bf k}]}}}{\Omega}\right)
+ \epsilon_{[{\bf k}]} \eeq
corresponding to eigenvectors
\beq
|\xi^{[{\bf{k}}]}_{[{\bf{n}}],m}\rangle &=& |\Phi_{[{\bf{n}}]}\rangle \otimes |\chi^{[\bf
k]}_{m}\rangle
\otimes |\Psi_{[{\bf{k}}]}\rangle.
\eeq
One can now re-express the Hamiltonians $H_i,\, i=1,\,2$  on  the separable
Hilbert space $\mathfrak H_{1} \otimes \mathfrak H_{2}$  spanned by the vectors $\left\{|\xi^{[{\bf k}]}_{[{\bf n}],m}\rangle, 
[{\bf k}] \in \Gamma \right \}_{[{\bf n}],m=0}^{\infty}$ as follows: \beq
H_1 &=& \Omega_{N} \sum_{[{\bf k}]\in \Gamma}\sum_{[{\bf{n}}],m=0}^{\infty} \mathcal E_{[{\bf{n}}]} \, |\xi^{[{\bf{k}}]}_{[{\bf{n}}],m}\rangle 
\langle \xi^{[{\bf{k}}]}_{[{\bf{n}}],m}| \qquad \mathcal E_{[{\bf{n}}]} = \sum_{l=1}^{N}n_{l} \crcr
H'_2 &=& \Omega \sum_{[{\bf k}]\in \Gamma}\sum_{[{\bf{n}}],m=0}^{\infty} (\mathcal
E^{[{\bf{k}}]}_{m} - \mathcal
E^{[{\bf{k}}]}_{0})|\xi^{[{\bf{k}}]}_{[{\bf{n}}],m}\rangle \langle
\xi^{[{\bf{k}}]}_{[{\bf{n}}],m}| \qquad \mathcal
E^{[{\bf{k}}]}_{m} - \mathcal
E^{[{\bf{k}}]}_{0} = m \eeq
where $H'_{2}$ is the shifted Hamiltonian obtained from $H_{2}$.

By analogy with the construction  (\ref{i}), the {\it multidimensional} coherent states
\cite{gazeau-novaes} generated by  the resulting  shifted Hamiltonian
$H' = H_{1} \otimes I_{\mathfrak H_{2}} + I_{\mathfrak H_{1}} \otimes H'_{2}$, by setting the summations over $[{\bf n}]$
 and $m$ independent, can be defined by \beq{\label{vcs01}} 
\fl |{\bf J},{\boldsymbol{\gamma}} ;J'_{[\bf
k]}, \gamma'_{[\bf
k]}\rangle &=& (\mathcal N_{1}({\bf
J}))^{-1/2}(\mathcal N_{2}(J'_{[\bf k]}))^{-1/2} \sum_{{[\bf n]}=0}^{\infty} \frac{{\bf
J}^{{\bf{n}}/2}e^{-\imath {\boldsymbol{\varepsilon}}_{[\bf n]}{\boldsymbol{\gamma}}}}{\sqrt{{\bf n}
!}}
\sum_{m= 0}^{\infty} \frac{J'^{m/2}_{[\bf
k]}e^{-\imath m\gamma'_{[\bf
k]}}}{\sqrt{m!}} \crcr  
&& \times |\xi^{[{\bf{k}}]}_{[{\bf{n}}],m}\rangle \eeq
where $J'_{[\bf k]} > 0$ and $\gamma'_{[\bf
k]} \in [0,2\pi)$, with
\beq
{\bf{J}}^{{\bf{n}}/2} := \prod_{l=1}^{N} J^{n_l/2}_{l} \qquad {\boldsymbol{\varepsilon}}_{[\bf n]}:=\left(
                              n_{1}, n_{2},  \cdots,
n_{N}
\right) \qquad {\boldsymbol{\gamma}} :=
                              \left(
                              \begin{array}{c}
                              \gamma_{1}\\
                               \gamma_{2} \\
                               \vdots \\
                                \gamma_{N}\\
                               \end{array}
                                \right)
\eeq

and the
 normalization condition \beq{\label{norm00}} \langle{\bf J},{\boldsymbol{\gamma}} ;J'_{[\bf
k]}, \gamma'_{[\bf
k]}|{\bf J},{\boldsymbol{\gamma}} ;J'_{[\bf
k]}, \gamma'_{[\bf
k]}\rangle = 1. \eeq
Since
\beq
H'|\xi^{[{\bf{k}}]}_{[{\bf{n}}],m}\rangle = (\Omega_{N}\mathcal E_{[{\bf{n}}]} + \Omega(\mathcal
E^{[{\bf{k}}]}_{m} - \mathcal
E^{[{\bf{k}}]}_{0}))|\xi^{[{\bf{k}}]}_{[{\bf{n}}],m}\rangle
\eeq
then we get
\beq
e^{-\imath H't}|\xi^{[{\bf{k}}]}_{[{\bf{n}}],m}\rangle = e^{-\imath \left(\Omega_{N}\mathcal E_{[{\bf{n}}]} + \Omega(\mathcal
E^{[{\bf{k}}]}_{m} - \mathcal
E^{[{\bf{k}}]}_{0})\right)t}|\xi^{[{\bf{k}}]}_{[{\bf{n}}],m}\rangle
\eeq
and thereby
\beq{\label{stab0009}}
e^{-\imath H't}|{\bf J},{\boldsymbol{\gamma}} ;J'_{[\bf
k]}, \gamma'_{[\bf
k]}\rangle = |{\bf J},{\boldsymbol{\gamma}} +
\Omega_{N}t\beta;J'_{[\bf k]},\gamma'_{[\bf
k]} + \Omega t\rangle \eeq
where $\beta$ is given as in (\ref{mat00}). Thus the states $|{\bf J},{\boldsymbol{\gamma}} ;J'_{[\bf
k]}, \gamma'_{[\bf
k]}\rangle$ are temporally stable.
They also satisfy the following action of the identity
\beq{\label{act00}}
&&\langle{\bf J},{\boldsymbol{\gamma}} ;J'_{[\bf
k]}, \gamma'_{[\bf
k]}|H'|{\bf J},{\boldsymbol{\gamma}} ;J'_{[\bf
k]}, \gamma'_{[\bf
k]}\rangle =\Omega_{N} \sum_{l=1}^{N} J_{l} + \Omega J'_{[\bf
k]}
\eeq
and the resolution of the identity on the Hilbert space $\mathfrak H_{1} \otimes \mathfrak H_{2}$ given by
\beq{\label{res00}}
&&\fl \sum_{[{\bf k}]\in \Gamma}\int_{\R^{N}_{+}}\int_{0}^{\infty}\int_{0}^{2\pi}\cdots \int_{0}^{2\pi}\int_{0}^{2\pi}|{\bf J},{\boldsymbol{\gamma}} 
;J'_{[\bf
k]}, \gamma'_{[\bf
k]}\rangle \langle{\bf J},{\boldsymbol{\gamma}} ;J'_{[\bf
k]}, \gamma'_{[\bf
k]}|  \mathcal N_{1}({\bf J}
)\mathcal N_{2}(J'_{[\bf k]}){\rm d}\mu(\gamma'_{[\bf k]}){\rm d}\rho({\boldsymbol{\gamma}}) \crcr
&& \fl \times  {\rm d}\nu({\bf J})
{\rm d}\varrho(J'_{[\bf k]})
= I_{\mathfrak H_{1} \otimes \mathfrak H_{2} }.
\eeq
The  moment problems
\beq
\frac{1}{n_{l} !}\int_{0}^{\infty}J^{n_{l}}_{l}d\nu_{l}(J_{l}) = 1 
\qquad \frac{1}{m !}\int_{0}^{\infty}J'^{m}_{[\bf k]}{\rm d}\varrho(J'_{[\bf k]}) = 1
\eeq
are explicitly solved with the  measures $d\nu_{l}(J_{l}) = e^{-J_{l}}dJ_{l}, 1\leq l\leq N$, and 
$d\varrho(J'_{[{\bf k}]}) = e^{-J'_{[{\bf k}]}}dJ'_{[{\bf k}]},$ respectively.
\subsubsection{Coherent states for eigenvalues with finite degeneracies.}{\label{paragph00}}
 We now deal  with the Hamiltonian $H$ possessing  eigenvalues with  finite degeneracies.
Specifically, when  $ \omega_{l}=\omega,$ for all $l,$ the eigenvalues  $E^{[{\bf k}]}_{[{\bf n}],m}$ become 
\beq E^{[{\bf k}]}_{n,m} = \omega n  + \left(\Omega m - \frac{g'^{2}_{\kappa_{[{\bf
k}]}}}{\Omega}\right) + \epsilon_{[{\bf k}]} \qquad n=n_1 + n_2 +\cdots  +n_N
\eeq corresponding to eigenvectors  given by \beq{\label{ti}}
|\xi^{[{\bf{k}}]}_{j-1,d(n)-j,m}\rangle  = |\Phi^{[{\bf k}]}_{j-1,d(n)-j}\rangle\otimes |\chi^{[\bf k]}_{m}\rangle
\otimes  |\Psi_{[{\bf
k}]}\rangle \qquad j=1,2,\dots ,d(n) \eeq where the vectors
$|\Phi^{[{\bf k}]}_{j-1,d(n)-j}\rangle$ are defined as in (\ref{exemlp00}), (\ref{exempl01}), (\ref{exempl02}) and (\ref{exempl03}). 
Since the eigenvalues
$E^{[{\bf k}]}_{n,m}, n=n_1+n_2+\cdots n_N,$ are degenerate
 ($\omega_{l}=\omega$ for all $l$), (with the degree of degeneracy   $ d(n) = C^{n}_{n+N-1}$),
then the eigenvalues of $ H_1$ generate the sequence  of
the quantities $(\epsilon_{{n}})_{{n \in \N}}$ given by (\ref{su})
 such that
 \beq{\label{ko}} E^{[{\bf{k}}]}_{n} &=&
\omega n. \eeq
Consequently, by similar
arguments as  in the previous section, we deduce  the shifted
Hamiltonians $H' = H_{1} \otimes I_{\hat \mathfrak H_{2}}+ I_{\hat \mathfrak H_{1}} \otimes H'_{2}$, on the separable Hilbert space $
\hat \mathfrak H_{1} \otimes \hat \mathfrak H_{2}$ spanned by the vectors
$|\xi^{[{\bf{k}}]}_{j-1,d(n)-j,m}\rangle $,  with: 
\beq && H_1 =
\omega\sum_{[{\bf k}]\in \Gamma}\sum_{m,n=0}^{\infty}\sum_{j=1}^{d(n)} n|\xi^{[{\bf{k}}]}_{j-1,d(n)-j,m}\rangle \langle
\xi^{[{\bf{k}}]}_{j-1,d(n)-j,m}|
\crcr
&& \fl H'_2 = \Omega\sum_{[{\bf k}]\in \Gamma}\sum_{m,n=0}^{\infty}\sum_{j=1}^{d(n)} (\mathcal
E^{[{\bf{k}}]}_{m} - \mathcal E^{[{\bf{k}}]}_{0}
)|\xi^{[{\bf{k}}]}_{j-1,d(n)-j,m}\rangle \langle
\xi^{[{\bf{k}}]}_{j-1,d(n)-j,m}|  \qquad  \mathcal
E^{[{\bf{k}}]}_{m} - \mathcal
E^{[{\bf{k}}]}_{0} = m.
\eeq
From (\ref{vcs00}) and (\ref{vcs01}) together by taking into account the independence of the summations over $n$ and $m$, respectively, 
the related coherent states for the Hamiltonian $H'$ can be defined by:
\beq{\label{vcsdiag}}
\fl |J,\gamma,\theta ;J'_{[\bf
k]}, \gamma'_{[\bf
k]}\rangle &=& (\mathcal N_{1}(J))^{-1/2}(\mathcal N_{2}(J'_{[\bf k]}))^{-1/2}  
\sum_{n=0}^{\infty} \sum_{j=1}^{d(n)}\frac{J^{n/2}e^{-\imath n\gamma}e^{-\imath j \theta}}{\sqrt{n
! d(n)}}
\sum_{m= 0}^{\infty} \frac{J'^{m/2}_{[\bf
k]}e^{-\imath m\gamma'_{[\bf
k]}}}{\sqrt{m!}}  \crcr
&& \times |\xi^{[{\bf{k}}]}_{j-1,d(n)-j,m}\rangle
\eeq
satisfying the normalization condition
\beq{\label{norm01}}
\langle J,\gamma,\theta ;J'_{[\bf
k]}, \gamma'_{[\bf
k]}|J,\gamma,\theta ;J'_{[\bf
k]}, \gamma'_{[\bf
k]} \rangle = 1
\eeq
and  the required properties of temporal stability, action identity and resolution of the identity on 
$\hat \mathfrak H_{1} \otimes \hat \mathfrak H_{2}$, respectively,
 as follows:
\beq{\label{stab01}}
e^{-\imath H't}|J,\gamma,\theta ;J'_{[\bf
k]}, \gamma'_{[\bf
k]}\rangle = |J,\gamma  + \omega t,\theta;J'_{[\bf
k]}, \gamma'_{[\bf
k]} + \Omega t\rangle
\eeq
\beq{\label{act01}}
\langle J,\gamma,\theta ;J'_{[\bf
k]}, \gamma'_{[\bf
k]}|H'|J,\gamma,\theta ;J'_{[\bf
k]}, \gamma'_{[\bf
k]}\rangle  = \omega J + \Omega J'_{[{\bf k}]}
\eeq
\beq{\label{res01}}
&& \fl \frac{1}{2\pi}\sum_{[{\bf k}]\in \Gamma}\int_{0}^{\infty}\int_{0}^{\infty}\int_{0}^{2\pi}\int_{0}^{2\pi}
\int_{0}^{2\pi}|J,\gamma,\theta ;J'_{[\bf
k]}, \gamma'_{[\bf
k]}\rangle \langle J,\gamma,\theta ;J'_{[\bf
k]}, \gamma'_{[\bf
k]} |   \mathcal N_{1}(J
)\mathcal N_{2}(J'_{[\bf k]}) \crcr
&& \fl \times {\rm d}\theta {\rm d}\mu(\gamma'_{[\bf k]}){\rm d}\rho(\gamma)
{\rm d}\nu_{N}(J)
{\rm d}\varrho(J'_{[\bf k]})
= I_{\hat \mathfrak H_{1} \otimes \hat \mathfrak H_{2} }
\eeq
where the measures  $d\nu_{N}(J) = \left[\frac{e^{-J}J^{N-1}}{(N-1)!} + 
J^{N}\delta (J) \right]
dJ$ and $d\varrho(J'_{[{\bf k}]}) = e^{-J'_{[{\bf k}]}}dJ'_{[{\bf k}]}$ 
solve the following moment problems
\beq
\frac{1}{n !d(n)}\int_{0}^{\infty}J^{n}d\nu_{N}(J) = 1 \qquad \frac{1}{m !}\int_{0}^{\infty}J'^{m}_{[\bf k]}{\rm d}\varrho(J'_{[\bf k]}) = 1,
\eeq
respectively.

\subsection{Extra diagonal case}
Setting $x_{\alpha\alpha'}=1$ if $\alpha
\neq \alpha'$ and $x_{\alpha\alpha'}=0$ if
$\alpha = \alpha'$, the model is now reduced to the Hamiltonian  $H$  given by
\beq{\label{hamilton}} H = &&
\sum_{i=1}^{N}\omega_{i}a^{\dag}_{i}a_{i}+ \Omega b^{\dag}b +
\sum_{\alpha=1}^{M}\epsilon_{\alpha}c^{\dag}_{\alpha} c_{\alpha} - g' \sum_{{}^{\alpha,\alpha'=1}_{\alpha \neq 
\alpha'}}^{M}(b^{\dag} + b) c^{\dag}_{\alpha}c_{\alpha'}. \eeq
This Hamiltonian is almost analogous to
(\ref{haml1}), except for  the fourth term.  The
mode of the c.m.  encompasses the electron-phonon coupling
constants  defined for fermion operators with different indices
(extra diagonal case).
The fourth term can be rewritten as
\beq
-g' \sum_{{}^{\alpha,\alpha'=1}_{\alpha \neq \alpha'}}^{M}(b^{\dag} + b) c^{\dag}_{\alpha}c_{\alpha'}  
= -g'(b+b^{\dag})\sum_{j=1}^{M}\left(c^{\dag}_{j}c_{j+1} + c^{\dag}_{j+1}c_{j}\right).
\eeq
Now, by use of a discrete Fourier transformation
\beq{\label{op001}}
c_{j} = \frac{1}{\sqrt{M}}\sum_{l=1}^{M}\eta^{jl}\tilde c_{l}
\eeq
where $\eta = exp\left\{\frac{\imath 2\pi}{M} \right\}$, the Hamiltonian $H$ can be recast as follows:
\beq
H = &&
\sum_{i=1}^{N}\omega_{i}a^{\dag}_{i}a_{i}+ \Omega b^{\dag}b +
\sum_{l=1}^{M}\tilde \epsilon_{l}\tilde c^{\dag}_{l} \tilde c_{l} - 2g'(b^{\dag} + b) 
\sum_{l=1}^{M}\cos\left\{\frac{2\pi l}{M}\right\}\tilde c^{\dag}_{l} \tilde c_{l}.
\eeq

Thus, the Hamiltonian $\tilde H_{2}$ given by
\beq
\tilde H_{2} = \Omega b^{\dag}b +
\sum_{l=1}^{M}\tilde \epsilon_{l}\tilde c^{\dag}_{l} \tilde c_{l} - g'(b^{\dag} + b) \sum_{l=1}^{M}\lambda_{l}\tilde c^{\dag}_{l} \tilde c_{l}
\eeq
where $\lambda_{l} := 2\cos\left\{\frac{2\pi l}{M}\right\}, 1 \leq l \leq M$, can be treated in the same way as the Hamiltonian $H_{2}$ of the 
diagonal case.
Then, similarly to (\ref{op000}), the corresponding eigenvalue problem  can be formulated as follows:
\beq
\tilde H_{2}\varphi  = \left[\Omega\left(b-\frac{g'_{\mu_{[{\bf
k}]}}}{\Omega}\right)^{\dag}\left(b- \frac{g'_{\mu_{[{\bf
k}]}}}{\Omega}\right)-\frac{{g'}^{2}_{\mu_{[{\bf
k}]}}}{\Omega} + \tilde \epsilon_{[{\bf k}]}\right]\varphi
\eeq
where $\tilde \epsilon_{[{\bf k}]}:=\sum_{k_{l} \in[{\bf
k}]}k_{l}\tilde \epsilon_{l}, 1 \leq l\leq M$, and
\beq
\mu_{[{\bf k}]} = \frac{1}{2} \sum_{l=1}^{M}(1-\delta_{k_{l},0}+\delta_{k_{l},1})\lambda_{k_{l}}
\eeq
with  $g'\mu_{[{\bf k}]} := g'_{\mu_{[{\bf k}]}}$ and $\lambda_{k_{l}} = 2\cos\left\{\frac{2\pi k_{l}}{M}\right\}, k_{l} = 0, 1$.

Therefore,  the  operators
\beq
\mathcal B^{\dag} = e^{-\imath \sqrt{2}\frac{g'_{\mu_{[{\bf
k}]}}}{\Omega}p}b^{\dag}e^{\imath \sqrt{2}\frac{g'_{\mu_{[{\bf
k}]}}}{\Omega}p}=b^{\dag}-\frac{g'_{\mu_{[{\bf
k}]}}}{\Omega}\eeq defined so that $[b,\sqrt{2}p]=\imath$
 yield the eigenvectors 
\beq
|\zeta^{[\bf k]}_{m}\rangle = \frac{(\mathcal B^{\dag})^{m}}{\sqrt{m
!}}e^{-\imath \sqrt{2}\frac{g'_{\mu_{[{\bf
k}]}}}{\Omega}p}|0\rangle \qquad m=0,1,2,\dots.
\eeq
Finally, 
the eigenvalues of $H$  can be deduced to give \beq \tilde E^{[{\bf
k}]}_{[{\bf{n}}],m} = \sum_{l=1}^{N}\omega_l n_l  +  \left(\Omega m -
\frac{g'^{2}_{\mu_{[{\bf k}]}}}{\Omega}\right)
+ \tilde \epsilon_{[{\bf k}]} \eeq
 associated with the eigenvectors
\beq
|\chi^{[{\bf{k}}]}_{[{\bf{n}}],m}\rangle &=& |\Phi_{[{\bf{n}}]}\rangle \otimes |\zeta^{[\bf
k]}_{m}\rangle
\otimes |\tilde \Psi_{[{\bf{k}}]}\rangle.
\eeq

\subsubsection{Coherent states in the nondegenerate case.}
Following the same construction as developed in the diagonal case, one can immediately provide the expression of the shifted Hamiltonian $H'$ 
 in the Hilbert space
 $\tilde \mathfrak H_{1} \otimes \tilde \mathfrak H_{2}$, spanned by the eigenvectors $\left\{|\chi^{[{\bf k}]}_{[{\bf n}],m}\rangle, 
[{\bf k}] \in \Gamma \right \}_{[{\bf n}],m=0}^{\infty},$ under the form
$H' = H_{1} \otimes I_{\tilde \mathfrak H_{2}} + I_{\tilde \mathfrak H_{1}} \otimes \tilde H'_{2},$ where
\beq
&&  H_1 =  \Omega_{N} \sum_{[{\bf k}]\in \Gamma}\sum_{[{\bf{n}}],m=0}^{\infty} \mathcal E_{[{\bf{n}}]} \, |\chi^{[{\bf{k}}]}_{[{\bf{n}}],m}\rangle 
\langle \chi^{[{\bf{k}}]}_{[{\bf{n}}],m}|\qquad \mathcal E_{[{\bf{n}}]} = \sum_{l=1}^{N}n_{l} \crcr
&& \tilde H'_2 = \Omega\sum_{[{\bf k}]\in \Gamma}\sum_{[{\bf{n}}],m=0}^{\infty} (\mathcal
E^{[{\bf{k}}]}_{m} - \mathcal
E^{[{\bf{k}}]}_{0})|\chi^{[{\bf{k}}]}_{[{\bf{n}}],m}\rangle \langle
\chi^{[{\bf{k}}]}_{[{\bf{n}}],m}| \qquad \mathcal
E^{[{\bf{k}}]}_{m} - \mathcal
E^{[{\bf{k}}]}_{0} = m.
\eeq
In the same vein as in (\ref{vcs01}), the resulting coherent states  
\beq
\fl |{\bf J},{\boldsymbol{\gamma}} ;\tilde J'_{[\bf
k]}, \tilde \gamma'_{[\bf
k]}\rangle &=& (\mathcal N_{1}({\bf
J}))^{-1/2}(\mathcal N_{2}(\tilde J'_{[\bf k]}))^{-1/2}  \sum_{{[\bf n]}=0}^{\infty} \frac{{\bf
J}^{{\bf{n}}/2}e^{-\imath {\boldsymbol{\varepsilon}}_{[\bf n]}{\boldsymbol{\gamma}}}}{\sqrt{{\bf n}
!}}
\sum_{m= 0}^{\infty} \frac{\tilde J'^{m/2}_{[\bf
k]}e^{-\imath m\tilde \gamma'_{[\bf
k]}}}{\sqrt{m!}} \crcr 
&& \times |\chi^{[{\bf{k}}]}_{[{\bf{n}}],m}\rangle.
\eeq
 satisfy the required properties determined by similar relations as in (\ref{norm00}), (\ref{stab0009}), (\ref{act00}) and (\ref{res00}), respectively.
The degenerate case can be discussed by mimicking step by step the reasoning developed in \ref{paragph00}.

\section{Concluding remarks}

In this work, we have investigated in detail all relevant physical
Hamiltonians considered  in quantum optics
 and condensed matter physics  to describe the
nanoparticle dynamics in terms of a system of interacting bosons and
fermions.   For these systems, we have built  vector coherent states
that well satisfy required mathematical properties of continuity,
temporal stability, resolution of the identity and action identity. A generalization to
multidimensional vector coherent states for the nondegenerate
Hamiltonians  has been also discussed, including the interesting case of the  Hamiltonian
 introduced by Yang et al.
\cite{yang-geller-dennis} pertaining to the study of a mechanism for electronic
energy relaxation in nanocrystals.

This work can be further generalized replacing 
 the ordinary
Fock-Heisenberg oscillator algebra  by the Jannussis
{\it{ et al.}} (\cite{jannussis-brodimas}) and Man'ko {\it{et al}}
(\cite{man'ko}) $f$-deformed oscillator algebra. The latter is generated by
the set of operators $\{a,a^{\dag},N\}$ and a well
defined deformation structure function $f$ of the number operator $N, $ such that \beq
A^{-} &=& af(N), \qquad A^{\dag} = f(N)a^{\dag}, \qquad {N} = A^{\dag}A^{-} =
Nf^{2}(N) \crcr  
[A^{-},A^{\dag}] &=& \{N+1\} - \{N\}. \eeq Such a study will be
in the core of a forthcoming paper.

\ack The authors are grateful to the referees for their useful comments which allowed to improve the paper. 
This work is partially
supported by the Abdus Salam International Centre for Theoretical
Physics (ICTP, Trieste, Italy) through the
 OEA-ICMPA-\mbox{Prj-15}. The ICMPA is in partnership with
the Daniel Iagolnitzer Foundation (DIF), France.

\appendix


\renewcommand{\theequation}{\Alph{section}.\arabic{equation}}
\setcounter{equation}{0}

\section{Temporal stability (\ref{genstab})}


From the definition,  it follows that
\beq
e^{-\imath H'_{D} t}|\mathfrak J,\gamma; [{\bf k}]\rangle &=&(\mathcal N({\bf J_{[{\bf k}]}}))^{-1/2}\sum_{j=0}^{\infty}\frac
{\mathfrak J^{j/2}e^{-\imath  \varepsilon \gamma}}{\sqrt{\rho(j)}}e^{-\imath H'_{D} t}|\Xi_{[Nj];[{\bf k}]}\rangle \cr
                                             & = &(\mathcal N({\bf
J_{[{\bf k}]}}))^{-1/2}\sum_{j=0}^{\infty}\frac
{\mathfrak J^{j/2}e^{-\imath \varepsilon \gamma}}{\sqrt{\rho(j)}}
e^{-\imath \left(E^{[\bf k]}_{[Nj]}-E^{[\bf k]}_{[
N0]}\right) td_{[{\bf k}]}}|\Xi_{[Nj];[{\bf k}]}\rangle . \nonumber
\eeq
Since
\beq
\Omega\varepsilon_{[Nj]} \beta = \Omega\left(
                              \epsilon_{1j}, \epsilon_{2j},  \cdots, 
\epsilon_{Nj}
\right) \left(
                      \begin{array}{c}
                         1 \\
                        \vdots \\
                         1 \\
                        \vdots \\
                        1 \\
                       \end{array}
                        \right)  = \Omega\sum_{l=1}^{N}\epsilon_{lj} = E^{[\bf k]}_{[Nj]}-E^{[\bf k]}_{[
N0]}
\eeq
we get
\beq
e^{-\imath H'_{D} t}|\mathfrak J,\gamma; [{\bf k}]\rangle 
 &=& (\mathcal N({\bf
J_{[{\bf k}]}}))^{-1/2}\sum_{j=0}^{\infty}\frac
{\mathfrak J^{j/2}
e^{-\imath (\varepsilon \gamma+ \Omega t \varepsilon_{[Nj]} \beta d_{[{\bf k}]})}}{\sqrt{\rho(j)}}|\Xi_{[Nj];[{\bf k}]}\rangle \cr
 &=&(\mathcal N({\bf
J_{[{\bf k}]}}))^{-1/2}\sum_{j=0}^{\infty}\frac
{\mathfrak J^{j/2}e^{-\imath \varepsilon(\gamma + \Omega t \beta  d_{[{\bf k}]})}}{\sqrt{\rho(j)}}|\Xi_{[Nj];[{\bf k}]}\rangle  \cr
 &=& |\mathfrak J,\gamma + \Omega t
\beta d_{[{\bf k}]}; [{\bf k}]\rangle.
\eeq

$\hfill{\square}$

\section{Action identity (\ref{genident})}


Using 
\beq
|\mathfrak J,\gamma;[{\bf k}]\rangle =   \left(
                      \begin{array}{c}
                         0 \\
                         0\\
                        \vdots \\
                         0 \\
                         |{\bf J}_{[{\bf k}]},\gamma_{[{\bf k}]}\rangle \\
                        0 \\
                        \vdots \\
                        0 \\
                        0\\
                       \end{array}
                        \right)  \quad \mbox{and}
\eeq
\beq
 H'_{D} = \left(
\begin{array}{ccccccc}
\tilde H'_{(0,\cdots,0,0)} & 0 & \cdots & \cdots & 0 & \cdots & 0\\
0 & \tilde H'_{(1,\cdots,0,0)} &  0 & \cdots & \cdots & \cdots  &  \vdots\\
\vdots & 0 & \cdots & 0 & \cdots & \cdots & \vdots\\
\vdots & \cdots & 0 & \tilde H'_{[{\bf k}]} & 0 & \cdots & \vdots \\
\vdots & \cdots & \cdots & 0 & \cdots & 0 & \vdots \\
\vdots & \cdots & \cdots & \cdots & 0 & \cdots & 0 \\
0 & \cdots & 0 & \cdots & \cdots & 0 & \tilde H'_{(1,\cdots,1,1)} \\
\end{array}
\right)
\eeq
we get
\beq
\langle \mathfrak J,\gamma;[{\bf k}]|H'_{D}|\mathfrak J,\gamma;[{\bf k}]\rangle  = 
  \langle{\bf J}_{[{\bf k}]},\gamma_{[{\bf k}]} |\tilde H'_{[{\bf k}]}|{\bf J}_{[{\bf k}]},\gamma_{[{\bf k}]}\rangle.
\eeq
With 

\beq
&&\tilde H'_{[{\bf k}]}|{\bf J}_{[{\bf k}]},\gamma_{[{\bf k}]}\rangle \cr
&=& 
\Omega \frac{1}{[\mathcal N({\bf J}_{[{\bf k}]})]^{1/2}}
\sum_{j= 0}^{\infty}\prod_{l=1}^{N} \frac{J^{j/2}_{[{\bf
k}]_l}e^{-\imath \epsilon_{lj}\gamma_{[{\bf k}]_l}}}{\sqrt{\rho_{l}(j)}}\left\{
\sum_{l=1}^{N}\epsilon_{lj}\right \} |\Psi_{[{\bf k}]} \otimes \Phi^{[{\bf k}]}_{[Nj]} \rangle
\eeq
we have
\beq
\fl&&\langle{\bf J}_{[{\bf k}]},\gamma_{[{\bf k}]} |\tilde H'_{[{\bf k}]}|{\bf J}_{[{\bf k}]},\gamma_{[{\bf k}]}\rangle \cr
\fl && = \Omega \frac{1}{\mathcal N({\bf J}_{[{\bf k}]})}
\sum_{j,p = 0}^{\infty}\prod_{l=1}^{N} \frac{J^{(j+p)/2}_{[{\bf
k}]_l}e^{-\imath (\epsilon_{lj}- \epsilon_{lp})\gamma_{[{\bf k}]_l}}}{\sqrt{\rho_{l}(j)\rho_{l}(p)}}\left\{
\sum_{l=1}^{N}\epsilon_{lj}\right \} \cr
&& \times  \langle  \Psi_{[{\bf k}]} \otimes \Phi^{[{\bf k}]}_{[Np]}|\Psi_{[{\bf k}]} \otimes \Phi^{[{\bf k}]}_{[Nj]} \rangle \cr
\fl&&=\Omega \frac{1}{\mathcal N({\bf J}_{[{\bf k}]})}
\sum_{j = 0}^{\infty}\prod_{l=1}^{N} \frac{J^{j}_{[{\bf
k}]_l}}{\rho_{l}(j)}\left\{
\sum_{l=1}^{N}\epsilon_{lj}\right \} \cr
\fl&& =  \Omega  \frac{1}{\mathcal N({\bf J}_{[{\bf k}]})}\sum_{j 
 =0}^{\infty} (\epsilon_{1j}\frac{J^{j}_{{[{\bf k}]}_1}
\cdots J^{j}_{{[{\bf k}]}_N}}{\rho_{1}(j)\cdots \rho_{N}(j)} + \epsilon_{2j}  \frac{J^{j}_{{[{\bf k}]}_1}
\cdots J^{j}_{{[{\bf k}]}_N}}{\rho_{1}(j)\cdots \rho_{N}(j)} \cr 
&&+ \cdots  + \epsilon_{Nj}\frac{J^{j}_{{[{\bf k}]}_1}
\cdots J^{j}_{{[{\bf k}]}_N}}{\rho_{1}(j)\cdots \rho_{N}(j)})\crcr
\fl&&= \Omega \frac{1}{\mathcal N({\bf J}_{[{\bf k}]})}\left\{J_{{[{\bf k}]}_1}\sum_{j
 =1}^{\infty}\frac{J^{j-1}_{{[{\bf k}]}_1}}{\prod_{q=1}^{j-1}\epsilon_{1q} }
\frac{J^{j}_{{[{\bf k}]}_2}
\cdots J^{j}_{{[{\bf k}]}_N}}{\rho_{2}(j)\cdots \rho_{N}(j)} \right\}\crcr
\fl&&+ \Omega \frac{1}{\mathcal N({\bf J}_{[{\bf k}]})}\left\{J_{{[{\bf k}]}_2}\sum_{j
 =1}^{\infty}\frac{J^{j-1}_{{[{\bf k}]}_2}}{\prod_{q=1}^{j-1}\epsilon_{2q} } \frac{J^{j}_{{[{\bf k}]}_1}J^{j}_{{[{\bf k}]}_3}\cdots 
J^{j}_{{[{\bf k}]}_N}}{\rho_{1}(j)\rho_{3}(j)\cdots \rho_{N}(j)} + \cdots\right \}\crcr
\fl&& + \Omega \frac{1}{\mathcal N({\bf J}_{[{\bf k}]})}\left\{\cdots + J_{{[{\bf k}]}_{N}}\sum_{j
 =1}^{\infty}\frac{J^{j-1}_{{[{\bf k}]}_N}}{\prod_{q=1}^{j-1}\epsilon_{Nq}}
 \frac{J^{j}_{{[{\bf k}]}_1}J^{j}_{{[{\bf k}]}_2}\cdots 
J^{j}_{{[{\bf k}]}_{N-1}}}{\rho_{1}(j)\rho_{2}(j)\cdots \rho_{N-1}(j)}\right \}\crcr
\fl&&= \Omega \frac{1}{\mathcal N({\bf J}_{[{\bf k}]})}\left\{J_{{[{\bf k}]}_1}\prod_{l=1}^{N} \sum_{j=0}^{\infty}
\frac{J^{j}_{[{\bf
k}]_l}}{\rho_{l}(j)} + \cdots + J_{{[{\bf k}]}_{N}}\prod_{l=1}^{N} \sum_{j=0}^{\infty}
\frac{J^{j}_{[{\bf
k}]_l}}{\rho_{l}(j)} \right\}\crcr
\fl&&= \Omega (J_{{[{\bf k}]}_1} + J_{{[{\bf k}]}_2} + \cdots +J_{{[{\bf k}]}_N}) = \Omega \sum_{l=1}^{N}J_{{[{\bf k}]}_l}.
\eeq
$\hfill{\square}$

\section{Resolutions of the identity (\ref{genres00}) and  (\ref{genres01})}


Let us consider the following quantities

\beq
{\bf{J}}_{[\bf{k}]} := \prod_{l=1}^{N} J_{[{\bf k}]_l} \qquad  \qquad
{\boldsymbol{\varepsilon}}_{[N_{j}]}:=\left(
                              \epsilon_{1j}, \epsilon_{2j},  \cdots,
\epsilon_{Nj}
\right)
\eeq
 such that
\beq
 {\boldsymbol{\varepsilon}}_{[N_{j}]} \cdot {\boldsymbol{\gamma}}_{[\bf{k}]}  
= \sum_{l=1}^{N}\epsilon_{lj} \gamma_{[{\bf k}]_l}.
\eeq

From the definition, we get 
\beq
\fl&& |{\bf J}_{[{\bf k}]},\gamma_{[{\bf k}]}\rangle  \langle{\bf J}_{[{\bf k}]},\gamma_{[{\bf k}]}| \crcr
\fl&&= \frac{1}{\mathcal N({\bf J}_{[{\bf k}]})}
\sum_{j,p = 0}^{\infty}\prod_{l=1}^{N} \frac{J^{(j+p)/2}_{[{\bf
k}]_l}e^{-\imath (\epsilon_{lj}- \epsilon_{lp})\gamma_{[{\bf k}]_l}}}{\sqrt{\rho_{l}(j)\rho_{l}(p)}} 
| \Psi_{[{\bf k}]} \otimes \Phi^{[{\bf k}]}_{[Np]}\rangle \langle \Psi_{[{\bf k}]} \otimes \Phi^{[{\bf k}]}_{[Nj]} |
\eeq
where   $\mathcal N({\bf J}_{[\bf k]}) =
\prod_{l=1}^{N} \mathcal N(J_{[{\bf k}]_l})$ and  $|\Phi^{[{\bf k}]}_{[N_j]}\rangle = 
\bigotimes_{l=1}^{N}|\Phi^{[{\bf k}]}_{lj}\rangle$. 

Then, it follows that
\beq
\fl&&\int_{0}^{L_1}\int_{0}^{L_2}
\cdots \int_{0}^{L_N} \left[\int_{\R}\int_{\R}\cdots \int_{\R}
|{\bf J}_{[\bf k]},{\boldsymbol{\gamma}}_{[\bf k]}\rangle \langle {\bf J}_{[\bf k]},{\boldsymbol{\gamma}}_{[\bf k]}|
 \mathcal N({\bf{J_{[{\bf k}]}}})d\mu_{B}({\boldsymbol{\gamma}}_{[\bf k]})\right]d\nu_{[{\bf k}]}({\bf J}_{[\bf k]})\crcr
\fl&&=  \sum_{j,p = 0}^{\infty}
\frac{| \Psi_{[{\bf k}]} \otimes \Phi^{[{\bf k}]}_{[Np]}\rangle \langle \Psi_{[{\bf k}]} 
\otimes \Phi^{[{\bf k}]}_{[Nj]} |}{\sqrt{\rho_{1}(j)\rho_{1}(p)}}
\int_{0}^{L_1}J^{\frac{j+p}{2}}_{[{\bf k}]_{1}}d\nu_{1}(J_{[{\bf k}]_{1}})\times 
\int_{\R}e^{-\imath (\epsilon_{1j}- \epsilon_{1p})\gamma_{[{\bf k}]_1}}d\mu_B(\gamma_{[{\bf k}]_{1}})\times \crcr
\fl&&\times \frac{1}
{\sqrt{\rho_{2}(j)\rho_{2}(p)}}\int_{0}^{L_2}J^{\frac{j+p}{2}}_{[{\bf k}]_{2}}d\nu_{2}(J_{[{\bf k}]_{2}}) \times 
\int_{\R}e^{-\imath (\epsilon_{2j}- \epsilon_{2p})\gamma_{[{\bf k}]_{2}}}d\mu_B(\gamma_{[{\bf k}]_{2}}) \times  \cdots \crcr
\fl&&\cdots \times   \frac{1}{\sqrt{\rho_{N}(j)\rho_{N}(p)}}
\int_{0}^{L_N}J^{\frac{j+p}{2}}_{[{\bf k}]_{N}}d\nu_{N}(J_{[{\bf k}]_{N}}) \times 
\int_{\R}e^{-\imath (\epsilon_{Nj}- \epsilon_{Np})\gamma_{[{\bf k}]_{N}}}d\mu_B(\gamma_{[{\bf k}]_{N}}).
\eeq

Since 

\beq
 \int_{\R} e^{\imath(\epsilon_{lp} - \epsilon_{lj})\gamma_{[{\bf k}]_{l}}}d\mu_B(\gamma_{[{\bf k}]_{l}}) = \cases{
              \begin{array}{lll}
             0 \qquad \mbox{if} \qquad p\neq j \\
               \\
             1 \qquad \mbox{if}  \qquad p = j
               \end{array}}
\eeq

then

\beq
\fl&&\int_{0}^{L_1}\int_{0}^{L_2}
\cdots \int_{0}^{L_N} \left[\int_{\R}\int_{\R}\cdots \int_{\R}
|{\bf J}_{[\bf k]},{\boldsymbol{\gamma}}_{[\bf k]}\rangle \langle {\bf J}_{[\bf k]},{\boldsymbol{\gamma}}_{[\bf k]}|
 \mathcal N({\bf{J_{[{\bf k}]}}})d\mu_{B}({\boldsymbol{\gamma}}_{[\bf k]})\right]d\nu_{[{\bf k}]}({\bf J}_{[\bf k]})\crcr
\fl&& = 
\sum_{j= 0}^{\infty}
\frac{| \Psi_{[{\bf k}]} \otimes \Phi^{[{\bf k}]}_{[Nj]}\rangle \langle \Psi_{[{\bf k}]} \otimes \Phi^{[{\bf k}]}_{[Nj]} |}{\rho_{1} (j)}
\int_{0}^{L_1}J^{j}_{[{\bf k}]_{1}}d\nu_{1}(J_{[{\bf k}]_{1}})\times   
\frac{1}{\rho_2(j)}
\int_{0}^{L_2}J^{j}_{[{\bf k}]_{2}}d\nu_{2}(J_{[{\bf k}]_{2}})   \crcr
\fl&&\times  \cdots \times   \frac{1}{\rho_N(j)}\int_{0}^{L_N}J^{j}_{[{\bf k}]_{N}}d\nu_{N}(J_{[{\bf k}]_{N}}) \crcr
\fl&&=   \sum_{j=0}^{\infty}
| \Psi_{[{\bf k}]} \otimes \Phi^{[{\bf k}]}_{[Nj]}\rangle \langle \Psi_{[{\bf k}]} \otimes \Phi^{[{\bf k}]}_{[Nj]} |
\eeq

where the  measures $d\nu_{l}(J_{[{\bf k}]_{l}}), 1\leq l\leq N$ 
solve the  moment problems
\beq{\label{line00}}
\frac{1}{\rho_{l}(j)}\int_{0}^{L_{l}}J^{j}_{[{\bf k}]_{l}}d\nu_{l}(J_{[{\bf k}]_{l}}) = 1 \qquad 
\int_{0}^{L_{l}}d\nu_{l}(J_{[{\bf k}]_{l}}) = 1.
\eeq


We have 
\beq{\label{linec}}
\fl&&\int_{\mathcal D^{2^{M}}}|\mathfrak J,{\boldsymbol{\gamma}};[{\bf k}]\rangle 
\langle \mathfrak J,{\boldsymbol{\gamma}};[{\bf k}]| \mathcal N({\bf{J_{[{\bf k}]}}})d\mu_{B}({\boldsymbol{\gamma}})d\nu(\mathfrak J)\crcr
\fl&&= \sum_{{j,p}=0}^{\infty}\int_{\mathcal D}d\mu_B(\gamma_{(0,\cdots,0,0)})
d\nu_{(0,\cdots,0,0)}({\bf J}_{(0,\cdots,0,0)})\int_{\mathcal D}
d\mu_B(\gamma_{(1,\cdots,0,0)})d\nu_{(1,\cdots,0,0)}({\bf J}_{(1,\cdots,0,0)}) \crcr
\fl&& \cdots\int_{\mathcal D}d\mu_B(\gamma_{[{\bf k}]})d\nu_{[{\bf k}]}({\bf J}_{[{\bf k}]})\cdots 
\int_{\mathcal D}d\mu_B(\gamma_{(1,\cdots,1,1)})d\nu_{(1,\cdots,1,1)}({\bf J}_{(1,\cdots,1,1)})\times \crcr
\fl&& \times 
diag(\frac{{\bf J}^{(j+p)/2}_{(0,0, \dots, 0,0)}
e^{-\imath (\varepsilon_{[Nj]} - \varepsilon_{[Np]})\gamma_{(0,0, \dots, 0,0)}}}{\sqrt{\rho(j)\rho(p)}}, 
\frac{{\bf J}^{(j+p)/2}_{(1,0, \dots 0,0)}
e^{-\imath (\varepsilon_{[Nj]} - \varepsilon_{[Np]})\gamma_{(1,0, \dots, 0,0)}}}{\sqrt{\rho(j)\rho(p)}}, \cdots, \crcr
\fl&&\frac{{\bf J}^{(j+p)/2}_{[{\bf k}]}
e^{-\imath (\varepsilon_{[Nj]} - \varepsilon_{[Np]})\gamma_{[\bf k]}}}{\sqrt{\rho(j)\rho(p)}}, \cdots, 
\frac{{\bf J}^{(j+p)/2}_{(1,1, \dots, 1,1)}
e^{-\imath (\varepsilon_{[Nj]} - \varepsilon_{[Np]})\gamma_{(1,1, \dots, 1,1)}}}{\sqrt{\rho(j)\rho(p)}})
|\Psi_{[\bf k]}\otimes\Phi^{[\bf k]}_{[Nj]}\rangle
\langle\Psi_{[\bf k]}\otimes\Phi^{[\bf k]}_{[Np]}|\crcr
\fl&&= \sum_{{j}=0}^{\infty}\int_{0}^{L_1} \cdots \int_{0}^{L_N}d\nu_{(0,\cdots,0,0)}({\bf J}_{(0,\cdots,0,0)})
\int_{0}^{L_1} \cdots \int_{0}^{L_N} d\nu_{(1,\cdots,0,0)}({\bf J}_{(1,\cdots,0,0)}) \cr
\fl&&\cdots\int_{0}^{L_1} \cdots \int_{0}^{L_N}d\nu_{[{\bf k}]}({\bf J}_{[{\bf k}]})\cdots \times 
 \int_{0}^{L_1} \cdots \int_{0}^{L_N}d\nu_{(1,\cdots,1,1)}({\bf J}_{(1,\cdots,1,1)}) \cr
\fl &&\times
diag(\frac{{\bf J}^{j}_{(0,0, \dots, 0,0)}}{\rho(j)}, \frac{{\bf J}^{j}_{(1,0, \dots 0,0)}}{\rho(j)}, \cdots, 
\frac{{\bf J}^{j}_{[{\bf k}]}}{\rho(j)}, \cdots, \frac{{\bf J}^{j}_{(1,1, \dots, 1,1)}}{\rho(j)})
|\Psi_{[\bf k]}\otimes\Phi^{[\bf k]}_{[Nj]}\rangle
\langle\Psi_{[\bf k]}\otimes\Phi^{[\bf k]}_{[Nj]}|.
\eeq

From  (\ref{line00}) and (\ref{linec}), we get
\beq
\fl&&\sum_{[{\bf k}]\in \Gamma}\int_{\mathcal D^{2^{M}}}|\mathfrak J,{\boldsymbol{\gamma}};[{\bf k}]\rangle 
\langle \mathfrak J,{\boldsymbol{\gamma}};[{\bf k}]| \mathcal N({\bf{J_{[{\bf k}]}}})d\mu_{B}({\boldsymbol{\gamma}})d\nu(\mathfrak J)\crcr
\fl&&= \sum_{[{\bf k}]\in \Gamma}\sum_{{j}=0}^{\infty}|\Psi_{[\bf k]}\otimes\Phi^{[\bf k]}_{[Nj]}\rangle
\langle\Psi_{[\bf k]}\otimes\Phi^{[\bf k]}_{[Nj]}| = \mathbb I_{2^{M}} \otimes I_{\hat \mathfrak H}
\eeq

which ends the proof.

$\hfill{\square}$

\section{Resolution of the identity (\ref{res01})}


From the definition  (\ref{vcsdiag}), we get for a given $[\bf
k] \in \Gamma$
\beq
&&|J,\gamma,\theta ;J'_{[\bf
k]}, \gamma'_{[\bf
k]}\rangle \langle J,\gamma,\theta ;J'_{[\bf
k]}, \gamma'_{[\bf
k]}|\crcr
&&= \frac{1}{[\mathcal N_{1}(
J)\mathcal N_{2}(J'_{[\bf k]})]}\sum_{n,p =0}^{\infty}\sum_{l,m = 0}^{\infty}\sum_{j=1}^{d(n)}\sum_{q = 1}^{d(p)}
\frac{J^{\frac{n + p}{2}}e^{-\imath (n-p)\gamma}e^{\imath(q-j)\theta}}{\sqrt{n ! d(n)p ! d(p)}}\crcr
&&\times\frac{J'^{\frac{l+m}{2}}_{[\bf k]}e^{-\imath (m-l) \gamma'_{[\bf k]}}}{\sqrt{{l !m !}}}
 |\xi^{[{\bf{k}}]}_{j-1,n-j+1,m}\rangle\langle \xi^{[{\bf{k}}]}_{q-1,p-q+1,l}|
\eeq
with $|\xi^{[{\bf{k}}]}_{j-1,n-j+1,m}\rangle =|\Phi_{j-1,n-j+1}\rangle \otimes |\chi^{[{\bf k}]}_{m}\rangle \otimes 
|\Psi_{[{\bf k}]}\rangle$. 

Taking $d\rho(\gamma) = \frac{1}{2\pi}d\gamma, 
d\mu(\gamma'_{[{\bf k}]}) = \frac{1}{2\pi}d\gamma'_{[{\bf k}]}$, it follows that
\beq
&&\fl\int_{0}^{\infty}\int_{0}^{\infty}\int_{0}^{2\pi}\int_{0}^{2\pi}
\int_{0}^{2\pi}|J,\gamma,\theta ;J'_{[\bf
k]}, \gamma'_{[\bf
k]}\rangle \langle J,\gamma,\theta ;J'_{[\bf
k]}, \gamma'_{[\bf
k]} |  \cr
&& \fl\times  \, \mathcal N_{1}(J
)\mathcal N_{2}(J'_{[\bf k]}) {\rm d}\theta  {\rm d}\mu(\gamma'_{[\bf k]}){\rm d}\rho(\gamma)
{\rm d}\nu_{N}(J)
{\rm d}\varrho(J'_{[\bf k]})\crcr
&&\fl =    \sum_{n,p =0}^{\infty}\sum_{l,m = 0}^{\infty}\sum_{j=1}^{d(n)}\sum_{q = 1}^{d(p)}
\frac{ |\xi^{[{\bf{k}}]}_{j-1,n-j+1,m}\rangle\langle \xi^{[{\bf{k}}]}_{q-1,p-q+1,l}|}{\sqrt{n ! d(n)p ! d(p)}}\int_{0}^{\infty}J^{\frac{n + p}{2}}
d\nu_{N}(J)\times \frac{1}{2\pi} \int_{0}^{2\pi}e^{-\imath (n-p)\gamma}{\rm d}\gamma \crcr
&& \fl\times \int_{0}^{2\pi} e^{\imath(q-j)\theta}{\rm d}\theta \times  
\frac{1}{\sqrt{m ! l!}}\int_{0}^{\infty}J'^{\frac{l+m}{2}}_{[\bf k]}{\rm d}\varrho(J'_{[\bf k]})\times
\frac{1}{2\pi}\int_{0}^{2\pi}e^{-\imath (m-l) \gamma'_{[\bf k]}}{\rm d}\gamma'_{[\bf k]}.
\eeq

Using the fact

 \beq
 \int_{0}^{2\pi} e^{\imath(q-j)\theta}{\rm d}\theta = \cases{
              \begin{array}{lll}
              0 \qquad \mbox{if} \qquad q \neq j \\
               \\
             2\pi \qquad \mbox{if}  \qquad q=j
               \end{array}}
\eeq

we introduce the factor  $1/2\pi$ such that 

\beq{\label{partresol}}
&&\fl\frac{1}{2\pi}\int_{0}^{\infty}\int_{0}^{\infty}\int_{0}^{2\pi}\int_{0}^{2\pi}
\int_{0}^{2\pi}|J,\gamma,\theta ;J'_{[\bf
k]}, \gamma'_{[\bf
k]}\rangle \langle J,\gamma,\theta ;J'_{[\bf
k]}, \gamma'_{[\bf
k]} |  \cr
&& \fl\times  \, \mathcal N_{1}(J
)\mathcal N_{2}(J'_{[\bf k]}) {\rm d}\theta {\rm d}\mu(\gamma'_{[\bf k]}){\rm d}\rho(\gamma)
{\rm d}\nu_{N}(J)
{\rm d}\varrho(J'_{[\bf k]})\crcr
&& \fl = \sum_{n,p =0}^{\infty}\sum_{l,m = 0}^{\infty}\sum_{j=1}^{d(n)}\sum_{q = 1}^{d(p)}
\frac{|\xi^{[{\bf{k}}]}_{j-1,n-j+1,m}\rangle\langle \xi^{[{\bf{k}}]}_{q-1,p-q+1,l}|}{\sqrt{n ! d(n)p ! d(p)}}\int_{0}^{\infty}J^{\frac{n + p}{2}}
d\nu_{N}(J) \times \frac{1}{2\pi}\int_{0}^{2\pi}e^{-\imath (n-p)\gamma}{\rm d}\gamma \crcr
&& \fl\times  \frac{1}{2\pi} \int_{0}^{2\pi} e^{\imath(q-j)\theta}{\rm d}\theta \times  
\frac{1}{\sqrt{m ! l!}}\int_{0}^{\infty}J'^{\frac{l+m}{2}}_{[\bf k]}{\rm d}\varrho(J'_{[\bf k]})\times
\frac{1}{2\pi}\int_{0}^{2\pi}e^{-\imath (m-l) \gamma'_{[\bf k]}}{\rm d}\gamma'_{[\bf k]}\crcr
&& \fl = \sum_{n,p =0}^{\infty}\sum_{l,m = 0}^{\infty}\sum_{j=1}^{d(n)}\sum_{q = 1}^{d(p)} \delta_{qj}\delta_{np}\delta_{ml}
\frac{|\xi^{[{\bf{k}}]}_{j-1,n-j+1,m}\rangle\langle \xi^{[{\bf{k}}]}_{q-1,p-q+1,l}|}{\sqrt{n ! d(n)p ! d(p)}}\int_{0}^{\infty}J^{\frac{n + p}{2}}
d\nu_{N}(J)\crcr
&& \fl\times  
\frac{1}{\sqrt{m ! l!}}\int_{0}^{\infty}J'^{\frac{l+m}{2}}_{[\bf k]}{\rm d}\varrho(J'_{[\bf k]})
  \cr
&& \fl = \sum_{n=0}^{\infty}\sum_{m = 0}^{\infty}\sum_{j=1}^{d(n)}
\frac{|\xi^{[{\bf{k}}]}_{j-1,n-j+1,m}\rangle\langle \xi^{[{\bf{k}}]}_{j-1,n-j+1,m}|}{n ! d(n)}
\int_{0}^{\infty}J^{n}d\nu_{N}(J)\times  \frac{1}{m !}\int_{0}^{\infty}J'^{m}_{[\bf k]}{\rm d}\varrho(J'_{[\bf k]}) \crcr
&& \fl = \sum_{n=0}^{\infty}\sum_{m = 0}^{\infty}\sum_{j=1}^{d(n)}|\xi^{[{\bf{k}}]}_{j-1,n-j+1,m}\rangle\langle \xi^{[{\bf{k}}]}_{j-1,n-j+1,m}|
\eeq
where the measures $d\nu_{N}(J) = \left[\frac{e^{-J}J^{N-1}}{(N-1)!} + 
J^{N}\delta (J) \right]
dJ$ and $d\varrho(J'_{[{\bf k}]}) = e^{-J'_{[{\bf k}]}}dJ'_{[{\bf k}]}$ 
solve the following moment problems
\beq
\frac{1}{n !d(n)}\int_{0}^{\infty}J^{n}d\nu_{N}(J) = 1 \qquad \frac{1}{m !}\int_{0}^{\infty}J'^{m}_{[\bf k]}{\rm d}\varrho(J'_{[\bf k]}) 
= 1
\eeq
respectively.

Since the identity operator is provided on the Hilbert space $\hat \mathfrak H_{1} \otimes \hat \mathfrak H_{2}$ as

\beq{\label{ident}}
\sum_{[{\bf k}]\in \Gamma}\sum_{n=0}^{\infty}\sum_{m = 0}^{\infty}\sum_{j=1}^{d(n)}
|\xi^{[{\bf{k}}]}_{j-1,n-j+1,m}\rangle \langle \xi^{[{\bf{k}}]}_{j-1,n-j+1,m}|  = I_{\hat \mathfrak H_{1} \otimes \hat \mathfrak H_{2} }
\eeq

from the line (\ref{partresol}),  after a summation on all $[\bf k] \in \Gamma$, we obtain by use of (\ref{ident})

\beq
&&\frac{1}{2\pi}\sum_{[{\bf k}]\in \Gamma}\int_{0}^{\infty}\int_{0}^{\infty}\int_{0}^{2\pi}\int_{0}^{2\pi}
\int_{0}^{2\pi}|J,\gamma,\theta ;J'_{[\bf
k]}, \gamma'_{[\bf
k]}\rangle \langle J,\gamma,\theta ;J'_{[\bf
k]}, \gamma'_{[\bf
k]} |  \cr
&& \times  \, \mathcal N_{1}(J
)\mathcal N_{2}(J'_{[\bf k]}) {\rm d}\theta {\rm d}\mu(\gamma'_{[\bf k]}){\rm d}\rho(\gamma)
{\rm d}\nu_{N}(J)
{\rm d}\varrho(J'_{[\bf k]}) = I_{\hat \mathfrak H_{1} \otimes \hat \mathfrak H_{2} }
\eeq

which completes the proof.

$\hfill{\square}$

\end{document}